\documentclass[envcountsect]{svmult}
\pdfoutput=1
\usepackage[cmex10]{amsmath}

\usepackage{amssymb,amsfonts}

\smartqed

\usepackage{mathptmx}       % selects Times Roman as basic font
\usepackage{helvet}         % selects Helvetica as sans-serif font
\usepackage{courier}        % selects Courier as typewriter font
\usepackage{makeidx}         % allows index generation
\usepackage[pdftex]{graphicx}        % standard LaTeX graphics tool
\usepackage{multicol}        % used for the two-column index
\makeindex             % used for the subject index
\newcommand{\eps}{\ensuremath{\varepsilon}}

\newcommand{\ome}{\ensuremath{\omega}}

\newcommand{\nn}{\ensuremath{\nonumber}} 

\newcommand{\lp}{\ensuremath{\left (}}
\newcommand{\rp}{\ensuremath{\right )}}
\newcommand{\lbr}{\ensuremath{\left [}}
\newcommand{\rbr}{\ensuremath{\right ]}}

\newcommand{\eeq}{\end{equation}}
\newcommand{\beq}{\begin{equation}}

\newcommand{\mcD}{\ensuremath{\mathcal{D}}}

\newcommand{\mcG}{\ensuremath{\mathcal{G}}}

\newcommand{\mbI}{\ensuremath{\mathbf{I}}}
\newcommand{\mbV}{\ensuremath{\mathbf{V}}}
\newcommand{\mbU}{\ensuremath{\mathbf{U}}}

\newcommand{\mbAf}{\ensuremath{\mathbf{A}_{\mathrm{f}}}}
\newcommand{\mbAi}{\ensuremath{\mathbf{A}_{\infty}}}
\newcommand{\mbA}{\ensuremath{\mathbf{A}}}
\newcommand{\mbB}{\ensuremath{\mathbf{B}}}
\newcommand{\mbC}{\ensuremath{\mathbf{C}}}

\newcommand{\mbE}{\ensuremath{\mathbf{E}}}
\newcommand{\mbS}{\ensuremath{\mathbf{S}}}
\newcommand{\mbO}{\ensuremath{\mathbf{O}}}
\newcommand{\mbT}{\ensuremath{\mathbf{T}}}
\newcommand{\mbJ}{\ensuremath{\mathbf{J}}}

\newcommand{\mbW}{\ensuremath{\mathbf{W}}}
\newcommand{\mbP}{\ensuremath{\mathbf{P}}}

\newcommand{\WW}{\ensuremath{\mathbb{W}}}
\newcommand{\NN}{\ensuremath{\mathbb{N}}}
\newcommand{\RR}{\ensuremath{\mathbb{R}}}

\newcommand{\EX}{\ensuremath{\mathrm{E}}}

\newcommand{\mrH}{\ensuremath{\mathrm{H}}} 
\newcommand{\mrI}{\ensuremath{\mathrm{I}}} 
\newcommand{\OUT}{\ensuremath{\mathrm{OUT}}} 
\newcommand{\IN}{\ensuremath{\mathrm{IN}}} 
\newcommand{\Isup}{\ensuremath{\mathrm{I}^\infty}} 
\newcommand{\Hinf}{\ensuremath{\mathrm{H}_\infty}} 
\newcommand{\Hsup}{\ensuremath{\mathrm{H}^\infty}} 
\newcommand{\Iinf}{\ensuremath{\mathrm{I}_\infty}} 
\newcommand{\ARCS}{\ensuremath{\mathrm{ARCS}}} 
\newcommand{\Dom}{\ensuremath{\mathrm{SDOM}}} 

\newcommand{\out}{\ensuremath{\mathrm{out}}} 
\newcommand{\into}{\ensuremath{\mathrm{in}}} 

\newcommand{\mrO}{\ensuremath{\mathrm{O}}} 
 
\newcommand{\mrc}{\ensuremath{\mathrm{c}}} 
\newcommand{\dmbV}{\ensuremath{\partial\mathbf{V}}} 
\newcommand{\mbVo}{\ensuremath{\mathrm{int}\mathbf{V}}}

\begin{document}

\title*{Structural Routability of\\$n$-Pairs Information Networks%
\thanks{Published in {\em Information and Control in Networks} (eds. G. Como, B. Bernhardsson, A. Rantzer), Lecture Notes in Control and Information Sciences v. 40, Springer, pp. 215--239, 2014. 
This work was supported by Australian Research Council grant DP110102401. A preliminary version was presented in \cite{nairISIT11}.}
}
\titlerunning{G.N. Nair. Structural Routability of $n$-Pairs Information Networks}

\author{Girish N. Nair}
\authorrunning{G.N. Nair. Structural Routability of $n$-Pairs Information Networks}

\institute{Girish N. Nair \at Dept. Electrical and Electronic Engineering, University of Melbourne, VIC 3010, Australia,  \email{gnair@unimelb.edu.au}}

\maketitle

\abstract{Information does not generally behave like a conservative fluid flow in communication networks with multiple sources and sinks.
 However, it is often conceptually and practically useful to be able to associate separate data streams with each source-sink pair, 
with only routing and no coding performed at the network nodes. This raises the question of whether
 there is a nontrivial class of network topologies for which achievability is always equivalent to `routability', 
for any combination of source signals and positive channel capacities. This chapter considers  possibly cyclic, 
 directed, errorless networks with $n$ source-sink pairs and mutually independent source signals. 
%and a relaxed communication 
%objective  in terms of demanded  information rates at sinks. 
The concept of
 {\em downward dominance} is introduced and it is shown that, if the network topology is downward dominated,
 then the achievability of a given combination of source signals and channel capacities 
implies the existence of a feasible multicommodity flow.}

\section{Introduction}
\label{introsec}

In an  $n${\em-pairs} or {\em multiple unicast} communication network,  $n$ source signals must be conveyed 
to their corresponding sinks  without exceeding any channel capacities. 
Until quite recently, the belief 
was that this was possible 
iff there existed a  
{\em routing} solution, i.e. if every symbol generated by a source
could be carried without modification, over channels and through network nodes, until it reached
the sink. %%Each network node in this approach  simply transfers every incoming bit
%%onto a suitable outgoing channel, without changing the bit-value. 
At a macroscopic level, this is equivalent to
presuming the existence of a feasible {\em multicommodity flow} \cite{leightonJACM99}.

However, in  \cite{yeungTIT95,ahlswedeTIT00}, an example was constructed
of a 2-pairs communication network that did not admit a 
routing solution, but became admissible if nodes could 
perform modulo-2 arithmetic on incoming bits.
This counter-intuitive result started the field of {\em network coding}, in which nodes are permitted
to not just route incoming symbols, but also to perform causal functions on them,
so as to better exploit the network structure and the available channel capacities.       

It is now known that the capacity regions for $n$-pairs networks 
are not generally given by feasible multicommodity flows. 
%%although in \cite{yanTIT06},
%the capacity region of a  special class of 3-layered, acyclic $n$-pairs networks was
%%demonstrated to be given by  multicommodity flows.  
In \cite{adlerSIAM06}, $n$-pairs networks were constructed with coding capacity 
much larger than the routing capacity. Other related work includes  \cite{hanTIT11},
in which  a necessary and sufficient condition for broadcasting correlated sources
over erroneous channels was found, and 
 \cite{liTIT03}, in which linear network coding was shown to achieve capacity
for a multicast network.
 
Notwithstanding the power of network codes, routing/multicommodity flow solutions are appealing in several respects.  
Most obviously they are simpler, because network nodes   
are not required to  perform extra mathematical operations on arriving bits. 
In addition, because different data streams are not `hashed' together by means of some function, 
there is arguably less potential for cross-talk between different source-sink pairs,
arising for instance from nonidealities during implementation in the physical layer.
For similar reasons, routing may be preferred over network coding if security 
and privacy are important. 
Furthermore, being able to treat information as a conservative fluid flow
could potentially provide a simple basis to analyse communication requirements
in  areas outside  traditional multiterminal information theory,
e.g. networked feedback control and multi-agent coordination/consensus problems - see, e.g.
\cite{antsaklisSpecial07}.

These considerations raise the natural questions of whether there is a general class of 
network topologies 
on which achievability is always equivalent to the existence of a feasible multicommodity
flow.  This chapter aims to answer this questions for  
 possibly cyclic, 
 directed, errorless networks with $n$ source-sink pairs  and mutually independent source signals,
where  the goal is to reconstruct
source-signals perfectly at their respective sinks.
 The structural concept of
 {\em downward dominance} (Def. \ref{tridefn}) is introduced, 
and the main result (Thm. \ref{mainthm}) is that if the network topology is downward dominated
then the existence  of an achievable combination of source signals and channel capacities always implies
the existence  of a feasible multicommodity flow.

The proof relies on the iterative
construction of an {\em entropically feasible} multicommodity flow
(Def. \ref{inffeasdefn}).
%%Conversely, the existence of a feasible multicommodity flow 
As downward dominance
 inheres solely in the topology of the network,
this result suits 
situations where channels, switches, transceivers and interfaces are 
expensive to set up and difficult to move,
or where   
channel capacities and source-signal statistics
are unknown. On these structures, information can  always
be treated like a flow of conservative, immiscible fluids.

Downward  dominance is a more general
condition than the notion of `triangularisability' that was
introduced in the conference version \cite{nairISIT11} of this 
chapter. While it is not generally easy to verify in arbitrary
$n$-pairs networks, Lemmas \ref{simplelem} and \ref{simplerlem})
give simpler, sufficient conditions for it to hold.   
Several examples are then provided in Sect. \ref{exsec} to illustrate the applicability of
 Theorem \ref{mainthm} to various example networks, both cyclic and acyclic,
including but not confined to the directed cycles and lines studied in \cite{kramer06,harveyTIT06}.   

Although downward dominance is sufficient to guarantee 
that routing can always achieve the full coding capacity of a network,
%%capacity''any feasible combination of source signals, demand rates and channel capacities on a given topology,
it is not necessary,  and the  important question of finding  
a  more general - or even tight - structural condition   
remains open. In the concluding section, e potential directions for future work  are outlined. 

\subsection{Notation and Basic Terminology}    
\label{notsubsec}

For convenience,   
the basic notation and terminology used in this chapter are described below.
\begin{itemize}
\item The set of nonnegative integers (i.e. whole numbers) is denoted by $\WW$, the set of positive integers (i.e. natural numbers)
by $\NN$, and the set of positive reals by $\RR_{>0}$.
\item A contiguous set $\{i,i+1,\ldots , j\}$ of integers is denoted $[i:j]$.  
\item Other sets are usually written in boldface type.
\item Random variables (rv's) are written in upper case and their realisations are indicated in corresponding lower case.
\item The set operation $\mbA\setminus\mbB$ denotes $\mbA\cap\mbB^\mrc$. 
\item A discrete-time random signal or process  $\left (F(k)\right )_{k=0}^\infty$ is denoted $F$, 
and  the finite sequence $\left (F(k)\right )_{k=s}^t$ is 
denoted  $F(s:t)$.
%%,  with $F^{-1}$ being the empty sequence $\{\}$.
%% and $F\equiv F[0:\infty]$.
%%Lebesgue measure on $\RR^r$ is denoted $\lambda_r$ and 
%%\item  $\log_2^+(\cdot ):=\max\{0,\log_2 (\cdot )\}$.
\item Given a subscripted rv or signal $F_j$, with $j$ belonging to a countable set $\mbJ$,  
$F_\mbJ$ denotes the tuple  $(F_j)_{j\in\mbJ}$, arranged according to the order on $\mbJ$.  
\item The {\em entropy} of a discrete-valued rv $E$ is denoted $\mrH [E ]\geq 0$,
 %% and 
%%the {\em differential entropy} of a continuous-valued rv $F$ (with absolutely continuous distribution) is written $\mrh[F]\in\RR$. 
and the conditional entropy of $E$ given another rv $F$  is $\mrH [E|F ]:=\mrH[E,F]-\mrH[F]$.
\item The {\em mutual information} between rv's $E$ and $F$ is denoted 
$\mrI [E;F]:=\mrH [E ]-\mrH [E|F ]\geq 0$,  
%if $E$ is discrete-valued. 
%If $E$ is continuous-valued, H is replaced with the differential entropy h. 
and  the {\em conditional mutual information} between rv's $E$ and $F$ given $G$ is denoted
 $\mrI [E ; F | G]:=\mrH [E|G ]-\mrH [E|F,G ]$.
%% if $E$ is discrete-valued. If $E$ is continuous-valued, H is replaced with h.  
\item If $E$ and $F$ are random processes and $E$ is discrete-valued,  then 
the {\em entropy rates} of $E$, and
the {\em conditional entropy rate} of $E$ given (past and present) $F$  
are respectively defined as  
\begin{align*} 
\Hinf [E] &:=  \varliminf_{t\to\infty}\frac{\mrH [E(0:t) ]}{t+1},   
\\
\Hsup [E] &:=  \varlimsup_{t\to\infty}\frac{\mrH [E(0:t) ]}{t+1},  
\\
\Hinf [E|F] &:=  \varliminf_{t\to\infty}\frac{\mrH [E(0:t) | F(0:t) ]}{t+1},    
\end{align*} 
%%If $E$ is continuous-valued,  H is simply replaced with h.
\item If $E,F$ and $G$ are random processes,  then 
the {\em mutual information rates} of $E$ and $F$, and
the {\em conditional mutual information rate} of $E$ and $F$ given (past and present) $G$  
are respectively defined as  
\begin{align*} 
\Isup [E; F ] &:=  \varlimsup_{t\to\infty}\frac{\mrI [E(0:t) ; F(0:t) ]}{t+1},  
\\
\Iinf [E; F ] &:=  \varliminf_{t\to\infty}\frac{\mrI [E(0:t) ; F(0:t) ]}{t+1},   
%%label{defIinfty}
\\
\Iinf [E; F|G] &:=  \varliminf_{t\to\infty}\frac{\mrI [E(0:t) ; F(0:t) | G(0:t)]}{t+1}.  
%%\label{defIcinfty}
\end{align*} 
\item A {\em directed graph (digraph)} $(\mbV,\mbA)$ consists of a set $\mbV$ of {\em vertices}, 
and a set $\mbA$  of {\em arcs} that each represent a directed link between a particular pair of vertices.
\item The initial vertex of an arc is called its {\em tail} and the terminal vertex, its {\em  head}. 
\item  A {\em walk} in a digraph is an alternating sequence 
$\omega=(\nu_1,\alpha_1,\nu_2,\alpha_2,\ldots ,\alpha_k,\nu_{k+1})$, $k\geq 0$,  
of vertices and arcs, beginning and ending in vertices, 
s.t. each arc $\alpha_l$ connects the vertex $\nu_l$ to $\nu_{l+1}$.
Each vertex $\nu_j$ and arc $\alpha_l$ in the sequence  is said to {\em be in} the walk;
with a minor abuse of notation, this is denoted $\nu_j\in\ome$.  
\item A {\em path} is a walk with no loops, i.e. it passes through no vertex more than once, including the initial one.  
\item An {\em undirected path} is an alternating sequence 
$\omega=(\nu_1,\alpha_1,\nu_2,\alpha_2,\ldots ,\alpha_k,\nu_{k+1})$, $k\geq 0$,  
of vertices and arcs, beginning and ending in vertices, 
s.t. no vertex is repeated and each arc $\alpha_l$ connects the vertex $\nu_l$ to $\nu_{l+1}$, or  $\nu_{l+1}$ to $\nu_{l}$. 
\item A {\em cycle} is a walk in which the initial and final vertices are identical, but every other vertex occurs once. 
\item A {\em subpath}  of a path  $(\nu_1,\alpha_1,\nu_2,\alpha_2,\ldots ,\alpha_k,\nu_{k+1})$
   is a segment  $(\nu_l,\alpha_l,\nu_{l+1},\ldots ,\nu_{j})$ of it, where $1\leq l\leq j\leq k+1$.   
\item  A vertex $\nu$ is said to be {\em reachable} from another vertex $\mu$, 
 denoted  $\mu\leadsto\nu$, 
 if $\exists$ a path leading from 
  $\mu$ to $\nu$.
Equivalently, it is said that $\mu$ {\em can reach}  $\nu$.
The same terminology and notation apply, with analogous meaning, for pairs of arcs 
 as well as mixed pairs of arcs and vertices. 
E.g. given an arc $\beta$, $\mu\leadsto\beta$ means that there is a path from the vertex $\mu$ to the tail of $\beta$.
\item Similarly, a (vertex or arc )set $\mbW$ is said to be {\em reachable} from another set $\mbU$, 
denoted $\mbU\leadsto\mbW$, if there is an element of $\mbW$ that is reachable from 
an element of $\mbU$; equivalently, it is said that $\mbU$ {\em can reach} $\mbW$.  
\item For any vertex set $\mbU\subseteq\mbV$,  $\ARCS(\mbU)\subseteq\mbA$ is the set of arcs with tails in $\mbU$.  
\item  The notation $\OUT(\mbU)$ ($\IN(\mbU )$) represents the set of arcs in $\mbA$ that have tails (resp. heads) in a vertex set $\mbU\subseteq\mbV$ 
and heads (tails) $\in\mbV\setminus\mbU$.  If $\OUT(\mbU)$ ($\IN(\mbU )$) consists of a single arc,  this arc is denoted $\out(\mbU)$ ($\into(\mbU )$). 
When $\mbU$ is a singleton $\{\mu\}$, the braces are omitted.
\end{itemize}

\section{Problem Formulation}
\label{forsec}

A network of unidirectional, point-to-point channels 
may be modelled  using a digraph $(\mbV,\mbA)$,
where  the vertex set $\mbV$ represents %%network nodes, such as 
information sources, sinks, repeaters, routers etc.,
and the arc set $\mbA$ indicates 
%%vertex-to-vertex 
the directions of any channels between  nodes.
%%and $c\equiv (c_\alpha)_{\alpha\in\mbf{A}}\in\RR_{>0}^{|\mbA|}$ 
%%is a vector of positive and possibly infinite {\em arc capacities}. 
%%For simplicity, propagation delays in channels are neglected.\footnote{Note that a channel with a delay can
%%be explicitly incorporated in the framework here by   
%%representing it as a series of two arcs with a vertex in between, such that the signal on the second arc is a delayed version of that on the first.}
As usual with digraphs, it is assumed that no arc leaves and enters the same vertex, and that
at most one arc leads from the first to the second element of any given ordered pair of vertices. 
In other words,  every arc in $\mbA$ may be uniquely identified with a tuple $(\mu,\nu)\in\mbV^2$,
with $\mu\neq\nu$.\footnote{Such digraphs are sometimes called {\em simple}.} 
It is also assumed that the digraph is {\em connected}, i.e. there is an undirected path between any distinct pair of vertices.
%%To avoid trivial situations, it is also assumed that every vertex in $\mbV$ is {\em adjacent} to at least one other,
%%i.e. connected by an arc to or from another vertex.

In an $n$-pairs information network,  the locations of 
%%information 
sources and sinks
are respectively represented by disjoint sets  
$\mbS=\{\sigma_1,\ldots , \sigma_n\}$ and  
 $\mbT=\{\tau_1,\ldots,\tau_n\}$ of distinct vertices in $\mbV$,
with each 
%%information 
source $\sigma_i$ aiming to communicate to 
exactly one sink $\tau_i$. It is assumed that $\sigma_i\leadsto\tau_i$.
Let $\mbP$ denote the sequence $\lp (\sigma_i,\tau_i)\rp_{i=1}^n$
of source-sink pairs, arranged in a specified order.
%%In addition,
%%the locations of external noise sources are represented by 
%%a set $\mbN= \{\sigma_{n+1},\ldots ,\sigma_{n+m}\}\subset\mbV$ of $m\geq 0$ distinct vertices, where $\mbN\cap(\mbS\cup\mbT)=\emptyset$.
Without loss of generality, it is assumed that every source (sink) has 
no in-coming (resp. out-going) arcs and exactly one out-going (in-coming) arc. 
\footnote{If a source or sink were actually connected to multiple nodes in the network, 
it would be represented in the digraph by an auxiliary vertex connected by an arc (of infinite capacity) with a multiply-connected vertex.}
%%Further note that no source vertex can then be a sink vertex, i.e. $(\mbS\cup\mbN)\cap\mbT=\emptyset$.} 
The {\em boundary} $\dmbV$  of the network is the set $\mbS\cup\mbT$
of source and sink vertices, and its {\em interior} is $\mbVo:=\mbV\setminus\dmbV$.

%%every non-source vertex $\in\mbV\setminus\mbS\cup\mbN$ is reachable from some source vertex $\in\mbS\cup\mbN$.

Each channel in the network can transfer bits errorlessly up to a maximum average rate,
as specified by a positive {\em arc-capacity}  
$c_\alpha\in\RR_{>0}$. 
%%as defined by a vector  $c:= (c_\alpha)_{\alpha\in\mbf{A}}\in (0,\infty]^{|\mbA|}$  
%%associating  each arc $\alpha$ with a 
%%of positive {\em arc-capacities}. 
%%An arc is called {\em finite}  
%%{\em nondegenerate} 
%%if it has finite capacity;
In some situations, it may be natural to assign infinite capacity to certain arcs,\footnote{For instance, 
when a single network node  is represented as two `virtual' vertices connected by an arc
of unbounded capacity.} 
and the set of all such arcs is denoted $\mbAi\subset\mbA$. In particular, the arcs leaving sources are by convention assigned infinite capacity.
The set of   
finite-capacity arcs is written $\mbAf=\mbA\setminus\mbAi$, with associated arc-capacity vector $c:=(c_\alpha)_{\alpha\in\mbAf}\in\RR_{>0}^{|\mbAf|}$.  
The {\em structure} of the $n$-pairs information network  
%%The {\em input-output (IO) graph}
%%%of the $n$-pairs information network 
is defined as the tuple $\Sigma=(\mbV,\mbAf,\mbAi,\mbP)$.   
%%With noise sources identified,  $\Phi=(\mbV,\mbAf,\mbAi,\mbP,\mbN)$ is called the {\em full structure}.   

%%Define a reversed structure. Do we need $\mbN$? 

The communication signals in the network are
represented by a vector $S\equiv(S_\alpha)_{\alpha\in\mbA}$ of discrete-valued random processes called {\em arc signals}.
%%with $S_\alpha:\WW\to\RR^{m_\alpha}$ taken to be discrete-valued $\forall\alpha\in\mbA$. 
%%\equiv \left (S_\alpha(t)\right )_{t\in\WW}$ 
In particular, the arc signals leaving sources and entering sinks respectively represent the 
exogeneous inputs to and outputs from the network.
For convenience, the input signal $S_{\out(\sigma_i)}$ generated by the $i$-th source  $\sigma_i\in\mbS$ 
is called  $X_i$,    
and the output signal $S_{\into(\tau_i)}$  entering the $i$-th sink  $\tau_i\in\mbT$  
is called $Y_i$.
It is assumed throughout this chapter that the signals $X_1,\ldots, X_n$ are mutually independent processes
with strictly positive entropy rates $\Hinf[X_i]>0$. 
%%However, though each may itself be a correlated process.
%%, and that all signals 
%%on finite-capacity arcs 
%%are discrete-valued.

The arc-signal vector $S$ is assumed to have the following property:
\begin{definition}[Setwise Causality and Signal Graphs]
\label{causaldefn}
An arc-signal vector $S$ is called {\em setwise causal} on a structure
%%capacitated IO-graph  
$\Sigma=(\mbV,\mbAf,\mbAi,\mbP)$  
%%$(\mbV,\mbA,\mbP,\mbN,c)$  
%%tuple $(\mbV,\mbA,\mbS,\mbN,\mbT,c,S)$ is called an $n${\em-pairs capacitated signal-graph} 
%%it satisfies the following two properties:  
%%A capacitated signal-graph $(\mbV,\mbA,\mbS,\mbN,\mbT,c,S)$ is called {\em valid}  
%%if it meets the following two conditions: 
%%\item[V1.]  Each source-signal is statistically independent of the others. 
%%\footnote{However, it may itself be a correlated process.}
%%\begin{description}
%%\item[V1.] 
if all arc signals leaving  vertices in any internal vertex-set $\mbU\subseteq\mbVo$   
are causally determined by those entering $\mbU$ from outside it.
That is, $\forall \mbU\subseteq\mbVo$, 
 $\exists$ an operator  $g_\mbU$    
s.t. 
\beq
S_{\ARCS(\mbU)}(t) = g_\mbU\left (t, S_{\IN(\mbU)}(0:t) \right ), 
\ \ \ \forall t\in\WW,
%%   \ \ \mbox{(Set-Wise Causality)}. 
\label{causal}
\eeq
%%\item[V2.] The rate at which any arc signal 
%%conveys information about exogeneous inputs is bounded  by the arc-capacity,
%%\beq
%%\Iinf [S_\alpha;X]\leq c_\alpha,
%%\ \ \forall \alpha\in\mbA \ \ \mbox{(Arc-Capacity Bound)}. 
%%\label{validcapbnd}
%%\eeq    
%%The capacitated signal graph $(\mbV,\mbA,\mbP,\mbN,S,c)$ is 
%%then also called {\em valid}.
%%\end{description}
where $\ARCS(\mbU)\subseteq\mbA $ denotes the set of arcs leaving vertices of $\mbU$. 

The tuple  $(\Sigma,S)$ is then called a {\em signal graph}. 
 {\flushright $\Diamond$}
\end{definition}
%%\begin{center} 

%%\end{center}
%%If the  the arc-capacity vector $c$ It is called a {\em signal graph} when  is not given but statements A1--A2 hold, 
{\bf Remark:} Setwise causality is a strengthened version of the basic concept 
of {\em well-posedness} \cite{zhouBook} in feedback control theory. 
In a well-posed feedback system,
the current values of all internal and output signals  are uniquely determined
by the past and present values of external inputs.\footnote{In the linear, time-invariant context of \cite{zhouBook}, 
this is equivalent to  the corresponding transfer functions
being well-defined and proper.} Setwise causality
essentially imposes an analogous condition on any subcollection of nodes and associated signals, treated as a system.  
In  acyclic digraphs (i.e. in which every walk is a path), it is equivalent to 
causality at every internal vertex.
However,  feedback signals may be present in cyclic digraphs,  in which case vertex-wise causality 
cannot guarantee 
 (\ref{causal}) 
without further assumptions, 
%%on the vertex mappings 
e.g. a positive time-delay at every vertex.

%%The tuple $(\mbV,\mbA,\mbS,\mbN,\mbT,c,X,W)$ is called an {\em  information network} if these conditions are met.
%%Now, partition $\mbS$ so that the first $n\geq 1$ vertices $\sigma_1,\ldots ,\sigma_n$
%%represent information sources and the next $m\geq 0$ vertices, $\sigma_{n+1},\ldots ,\sigma_{n+m}$, 
%%represent  sources of any external interference or noise. 
%%In an $n${\em-pairs communication problem}, 

%%For notational convenience, denote the noise sources $\in\mbN$ as $\sigma_{n+1},\ldots,\sigma_{m+n}$.  

In the  $n$-pairs network problem studied here, the objective is 
for each sink to perfectly reconstruct
each source signal, block-by-block, using only causal operations 
and without exceeding any arc-capacities. 
%%In this chapter, the goal is to guarantee that the marginal information rates
%%supplied to the sinks exceed specified {\em demands}. 
This leads to the following definition:
\begin{definition}[Achievability] 
\label{achdefn}
Consider an  $n$-pairs information network with structure $\Sigma$,
 source-signal vector $X$ 
%%$\Gamma=(\mbV,\mbA,\mbP,\mbN,X,c)$ 
%%be a capacitated IO-graph for an $n$-pairs communication network,
and arc-capacity vector $c\in\RR_{>0}^{|\mbAf|}$. 
%%and  demand vector $d:=(d_i)_{i=1}^n\in\RR_{>0}^n$.
%%with $d_i \leq \Hinf[X_i]$, 
%%$\forall i\in [1:n]$.
The tuple $(\Sigma,X,c)$ is called {\em achievable}
  if $\exists$ a setwise-causal  arc-signal vector $S$ 
(Def. \ref{causaldefn})
and a positive integer $m\in\NN$ s.t.
\begin{align}
S_{\out(\sigma_i)} &= X_i, \ \ \forall i\in [1:n], 
\label{Xi}\\ 
%%\Iinf[Y_i;X_i] &\geq  d_i, \ \ \forall i\in [1:n],
Y_i(km-1) &= X_i\lp (k-1)m  :km-1\rp , \ \ \forall k\in\NN, i\in[1:n],
\label{demand}
\\
\Hsup [S_\alpha] &\leq  c_\alpha, 
\ \  \forall \alpha\in\mbAf. 
\label{capbnd}
\end{align}   
Such an $S$ is called a {\em solution} to the $n${\em-pairs information network problem} $(\Sigma,X,c)$.
The 
%%demand vector $d$ is then called achievable on $(\Sigma,X,c)$;  
 arc-capacity vector $c$ is called achievable on $(\Sigma,X)$ 
and $(X,c)$ is called  achievable on $\Sigma$.
{\flushright $\Diamond$} 
\end{definition}

{\bf Remarks:} This differs from standard definitions of network coding solutions in 
several  minor respects.
For instance, in \cite{ahlswedeTIT00,doughertyTIT05,kramer06,cannonsTIT06}  
and most of \cite{harveyTIT06},
the inequalities (\ref{capbnd}) are replaced by bounds
either on the cardinalities of channel alphabets, 
or on block-coding rates over a period of time.
In addition,  in previous formulations,  the sinks typically must reconstruct  the source signal
either perfectly and instantaneously  \cite{doughertyTIT05,cannonsTIT06,harveyTIT06}, 
which corresponds to setting $m=1$ in (\ref{demand}),
or else with arbitrarily small probability 
of decoding error over blocks of sufficiently large length $m$ \cite{ahlswedeTIT00,kramer06}. 

In this work,  bounds are imposed directly on entropies,
as in sec. VIII of \cite{harveyTIT06}, 
in order to  focus on the information-theoretic aspects of the problem.
Errorless reconstruction is demanded so as to  
enable the graphical characterisation of {\em informational dominance} from \cite{harveyTIT06}
to be used with very minor changes.  
However, perfect reconstruction is not required instantaneously in (\ref{demand}),
but only in blocks of length $m$.
This allows a solution $S$ to be interpreted 
operationally  in terms of   variable bit-rate   
 %%this is also essentially the case in \cite{doughertyTIT05,harveyTIT06,cannonsTIT06}.
codes.\footnote{In other words, if $S$ solves $(\Sigma,X,c)$,
then there exist variable bit-rate codes for each arc that
yield errorless, block-by-block reconstruction of the source-signals at their sinks,
with expected bit-rates at worst negligibly larger than arc-capacities.
Conversely,  if there exists a distributed entropy coding scheme 
that achieves perfect reconstruction of source-signals at their sinks in blocks of length  $m$,  
and with expected bit-rates no larger than the arc-capacities,
then this  yields a solution $S$ as  defined above.
However, these operational interpretations will not be
used in this article.}

Finally, it is conjectured that the results in this paper also apply if
(\ref{demand}) is relaxed so that $Y_i$ is causally determined by $X_i$, with
$\Hinf[Y_i]>0$.   

%%\footnote{In cite{doughertyTIT05,
%%cannonsTIT06},  
%% the left-hand side (LHS) of (\ref{capbnd}) is essentially replaced by 
%%the average bit-rate on that arc over a finite time-period; in \cite{  

%%with a specified set $\mbAfc$ of finite-capacity arcs. 
%%(where $Y_i$ is the arc-signal entering $\tau_i$).

%%The closure of the set of achievable demand vectors $d$ 
%%s.t. $(\Sigma)$ is achievable
%%is called 
%%the {\em demand region} $\mbD\subseteq\RR_{\geq 0}^n$ of $(\Sigma,X,c)$.  
%%For a given  
%an IO graph $(\mbV,\mbA,\mbP,\mbN)$ with constrained arc set $\mbA_\mrc$ and 
%%set $\mbAfc$ of finite-capacity  arcs, 
%%The closure of the set of achievable arc-capacity vectors 
%%$c$
%%$c'=(c_\alpha)_{\alpha\in\mbAfc}\in\RR_{>0}^{|\mbAfc|}$ 
%%s.t. $(\Sigma)$ is achievable 
%%is called 
%%the {\em arc-capacity region} 
%%$\mbC\subseteq\RR_{\geq 0}^{|\mbAf|}$
%%of $(\mbV,\mbA,\mbP,\mbN,\mbAfc)$.
%%$\mbC\subseteq\RR_{\geq 0}^{|\mbAfc|}$
%%of $(\Sigma,X)$. 

As mentioned in the introduction, 
%%the derivation of a finite set of conditions that completely characterises  achievability 
%%(essentially equivalent to determining the demand or capacity regions)  
%%for $n$-pairs networks is an open problem. 
it was once thought that 
a network was achievable\footnote{ignoring differences in the definition of achievability} 
iff it admitted a 
 routing solution. 
In the present context, this is equivalent  to
presuming the existence of an $(X,c)${\em-feasible multicommodity flow}, i.e.  
of a nonnegative tuple $f=(f_{\alpha,j})_{\alpha\in\mbA, j\in[1:n]}\in\RR^{|\mbA| n}_{\geq 0}$,
of bit-rates on each arc associated with every  source-sink pair,
s.t.
\begin{align}  
\sum_{j=1}^n f_{\alpha,j} 
&\leq  c_\alpha, \ \ \forall \alpha\in\mbAf
&& \text{(capacity bound)}; 
\label{mccapbnd}
\\
f_{\into(\tau_j),j}  
& = f_{\out(\sigma_j),j}= \Hinf [X_j ], \ \  
  \forall j\in [1:n] 
&& \text{(supply equals demand)},  
\label{mcdemand}
\\
%%f_{\out(\sigma_j),j}
%%& =  \Hinf [X_j ],  
%%\  \ \forall  j\in[1:n], 
%%&& \text{(sufficiency of supply)},  
%%\label{mcsupply} 
%%\\
\sum_{\alpha\in\IN(\nu)}f_{\alpha,j} 
&=   \sum_{\alpha\in\OUT(\nu)}f_{\alpha,j} 
 && \text{(conservation of flow)},
\label{mccons}
\end{align}
for any $j\in[1:n]$ and $\nu\in\mbV\setminus\left (\{\sigma_j\}\cup\{\tau_j\}\right )$.
Via an explicit counter-example, the article \cite{ahlswedeTIT00} showed that this intuitive notion was incorrect,
i.e. that although the existence of a feasible multicommodity flow is sufficient for achievability,
it is not generally necessary.  
This laid the foundations for {\em network coding}, in which nodes are permitted
to not just route incoming bits, but also to perform functions on them.
 
Nonetheless, routing/multicommodity-flow solutions have certain virtues, as discussed in Sect. \ref{introsec}.
This chapter poses the  question:
is there a general class of 
$n$-pairs information network structures $\Sigma$
in which the achievability of $(X,c)$ is  equivalent to the existence of an $(X,c)$-feasible multicommodity
flow $f$ (\ref{mccapbnd})--(\ref{mccons})?

Any $n$-pairs information network structure $\Sigma$ can support 
$(X,c)$-feasible multicommodity flows   
if the arc-capacities  are sufficiently larger than the source entropy rates,
provided each
sink is reachable from its source. 
However, there are examples of structures on which 
    an  $(X,c)$-feasible multicommodity flow does not exist if arc-capacities are reduced,
even though $(X,c)$ is still achievable (see Sect. \ref{exsec}).  

The aim of this chapter is to isolate certain structural properties that 
 ensure routability 
{\em over all achievable combinations of} $(X,c)$. 
Such properties would inhere solely in $\Sigma$,
suiting situations    
in which channels, switches, transceivers and interfaces are 
expensive to set up and difficult to move,
and/or where   
channel capacities and source-signal statistics
are variable or unknown.

%%to this question 
%%and the main result of this chapter, stated below, specifies graph-theoretic
%%conditions on $\Sigma$ that ensure fluidity.

\section{Preliminary Notions}
\label{prelimsec}

Before proceeding,   several 
existing graph-theoretic notions are needed.
Throughout this section, 
$\Sigma=(\mbV,\mbA,\mbP)\equiv (\mbV,\mbAf,\mbAi,\mbP)$ is the structure of an $n$-pairs information network
as described in  Sect. \ref{forsec}, 
and $\Gamma=(\Sigma,S)$ is its setwise-causal signal graph (Def. \ref{causaldefn}),
with source- and sink-signal vectors $X$ and $Y$. 

First, some largely familiar concepts are revisited.  
A path in an $n$-pairs information network that goes from a source $\sigma_i$ to its sink  $\tau_i$ 
is called  an  $i${\em-path}.  The set of all $i$-paths is called an $i${\em-bundle}, i.e.
the set of all acyclic walks via which information can be routed
from   $\sigma_i$ to $\tau_i$.
Given a set $\mbJ\subseteq[1:n]$, the set of all $i$-paths with $i\in\mbJ$ is called a $\mbJ${\em-bundle}
(not the same as the set of $\sigma_\mbJ\leadsto\tau_\mbJ$-paths, 
which contains it). 
Let $(\mbV^\mbJ,\mbA^\mbJ)$ denote the subgraph formed by all the vertices and arcs in the $\mbJ$-bundle.
In particular, $(\mbV^i,\mbA^i)$ is the subgraph formed by the $i$-bundle.
A vertex set $\mbU\subset\mbV^i$
such that $\sigma_i\in\mbU$ and $\tau_i\notin\mbU$
%%the arc set $\OUT(\mbU)$ 
is called an {\em $i$-cut}.
%%\footnote{In the literature, the term cut usually refers to the complementary pair $(\mbU,\mbUc)$,
%%but it is convenient here to assign it to 
%%the set of outgoing arcs, with 
%%$\mbU$.
%%possibly being nonunique.
%%} 

The  following concepts are adapted from \cite{harveyTIT06}, with minor changes in terminology.

\begin{definition}[Indirect $i$-Walks  --  Based on \cite{harveyTIT06}]
\label{indirectdefn}
An {\em indirect $i$-walk (i$i$-walk)} $\ome$ is an alternating  sequence $(\alpha_1,\beta_1,\ldots , \alpha_{j-1},\beta_{j-1},\alpha_j)$ of
forward- and reverse-oriented paths in the $n$-pairs structure $\Sigma$ such that
\begin{enumerate}
\item $\alpha_1$ begins with the $i$-th source vertex $\sigma_i$;  
\item both  $\alpha_\ell$ and $\beta_\ell$ end with the same vertex $\mu_\ell$,  $\forall\ell\in[1:j-1]$; 
\item  both $\beta_\ell$ and $\alpha_{\ell +1}$ begin from the same source vertex,  
 $\forall\ell\in[1:j-1]$;   
\item $\alpha_j$ ends with the sink vertex $\tau_i$; and  
\item every arc and vertex in $\ome$  can reach $\tau_i$. 
\end{enumerate}

An i$i$-walk $\ome$ is said to 
{\em bypass}  an arc-set $\mbC$ if
no arc in $\ome$ lies in $\mbC$. 
%%If no i$i$-walk bypasses $\mbC$, then  $\mbC$ is said to 
{\flushright $\Diamond$} 
\end{definition}

{\bf Remarks:} Note that the fifth condition above is equivalent to the requirement that
each joint vertex $\mu_\ell$ reaches $\tau_i$,  
$\forall\ell\in [1:j-1]$. 

An i$i$-walk as defined above is, in the terminology of \cite{harveyTIT06},
an {\em indirect walk} from $\out(\sigma_i)$
to $\into(\tau_i)$ in a subgraph $G(\emptyset,i)$.  
Similarly, an i$i$-walk that bypasses $\mbC$
is an indirect walk from $\out(\sigma_i)$
to $\into(\tau_i)$ in a subgraph $G(\mbC,i)$;
if such a bypass exists, then  $Y_i$ is not always fully determined by 
 $S_\mbC$, even  if all $i$-paths go through $\mbC$.    
See  Fig. 3 and Defs. 10 -- 11 {in} \cite{harveyTIT06}.

Indirect $i$-walks are related to the concept of 
{\em fd-separation} \cite{kramer06}.
In particular, if $S_\mbC$ fd-separates 
$X_i$ and $Y_i$ for any setwise causal $S$ (Def. \ref{causaldefn}), then all i$i$-walk's pass through $\mbC$; 
that is, an i$i$-walk that bypasses $\mbC$ corresponds to an undirected path between 
$X_i$ and $Y_i$  in a {\em functional dependence} subgraph
$\mcG_{X_i,S_{\mbC},Y_i}$ constructed according to the procedure in \cite{kramer06}.  

However, the converse is not generally true, i.e. `i$i$-separation'  
is a less stringent requirement. This is because  paths connecting 
$X_i$ and $Y_i$  in $\mcG_{X_i,S_{\mbC},Y_i}$ do not have to satisfy 
an analogue of the fifth condition, which arises from the requirement that  each sink 
 reproduce its source signal with perfect fidelity.
 For this to be possible, 
it turns out that each  joint vertex $\mu_\ell$ in an i$i$-walk must be able to reach  $\tau_i$.

Put another way,   requiring  $S_\mbC$ to fd-separate 
$X_i$ and $Y_i$ is equivalent to requiring 
that  {\em a)} $\mbC$ be an $i$-cut, 
and {\em b)} for each $j\neq i$, either all $\sigma_j\leadsto\tau_i$-paths (if any) bypass $\mbC$, 
or  all pass through it.   Under i$i$-separation,
 {\em (a)} must still hold, but  {\em (b)}  
is relaxed:  a source $\sigma_j$ can have  a path $\pi$ to $\tau_i$ 
that bypasses $\mbC$ as well as another  that passes through $\mbC$,
provided that $\pi$ is not the last leg of an i$i$-walk that bypasses   
$\mbC$.

\begin{definition}[Structural Dominance  -- Based on \cite{harveyTIT06}]
\label{infodomdefn}
For any arc-set $\mbB\subseteq\mbA$ in an $n$-pairs network, $\Dom(\mbB)$ 
is the smallest arc-set $\mbC\subseteq\mbA$ that satisfies the conditions below:
\begin{enumerate}
\item $\mbC\supseteq \mbB$
\item $\out(\sigma_i)\in \mbC$ iff $\into(\tau_i)\in \mbC$
\item If $\alpha\in\mbA$ is {\em downstream} from $\mbC$ 
-- i.e. all paths from sources to the tail of $\alpha$  pass through $\mbC$ --   
then $\alpha\in\mbC$.
\item If all indirect $i$-walks  (Def.\ref{indirectdefn}) pass through $\mbC$  then
$\out(\sigma_i),\into(\tau_i)\in\mbC$.
\end{enumerate}
The arcs in $\Dom(\mbB)$ are said to be {\em structurally dominated by} $\mbB$. 
{\flushright $\Diamond$} 
\end{definition}

{\em Remarks:} Note that $\Dom(\mbB)$ 
is the smallest such arc-set in the sense of being contained by every $\mbC\subseteq\mbA$ that satisfies criteria 1 -- 4.

 As noted in \cite{fragouliBook} (pp. 199--200),
$\Dom(\mbB)$ can be constructed by setting $\mbC=\mbB$,
 letting  $\mbT=\mbA\setminus\mbB\neq\emptyset$ be the set of arcs to be  tested,
 and then following this
greedy algorithm: 
\begin{description}[(iii)]
\item[(i)] Pick any arc $\alpha\in\mbT$.
\item[(ii)]  If  $\alpha$ satisfies any of the conditions 2 -- 4 {in} Def. \ref{infodomdefn},
 update $\mbC\leftarrow\mbC\cup\{\alpha\}$ and then $\mbT\leftarrow\mbA\setminus\mbC$;
else keep $\mbC$ the same and update  $\mbT\leftarrow\mbT\setminus\{\alpha\}$.
\item[(iii)]  If $\mbT=\emptyset$ then exit; else  go to step (i).
\end{description}
The final set $\mbC$ is then $\Dom(\mbB)$. However, 
the following lemma gives two quicker conditions for guaranteeing that a specific arc
lies in $\Dom(\mbB)$.

\begin{lemma}[Based on \cite{harveyTIT06}]
\label{sdomlem}
\begin{enumerate}
 \item If an arc $\alpha\in\mbA$  is downstream from $\mbB$, then $\alpha\in\Dom(\mbB)$.
\item If   all indirect $i$ walks (Def. \ref{indirectdefn}) pass through $\mbB$ 
then $\out(\sigma_i),\into(\tau_i)\in \Dom(\mbB)$. 
\end{enumerate}
\end{lemma} 
\begin{proof}
If either of these criteria hold, then  
the  relevant arcs -- $\alpha$,  $\out(\sigma_i)$, $\into(\tau_i)$ --  
must lie inside any arc-set $\mbC$ that satisfies the 1st to 4th conditions in Def. \ref{infodomdefn}.
As $\Dom(\mbB)$ is such a set, the lemma follows.  
\qed
\end{proof}

The significance of structural dominance arises from the following result:
\begin{theorem}[Informational Dominance -- Based on \cite{harveyTIT06}] 
\label{harveythm}
Consider any arc $\alpha\in\mbA$ and arc-set $\mbB\subseteq\mbA$ in an $n$-pairs network with structure $\Sigma$.   
If  $\alpha\in\Dom(\mbB)$ (Def. \ref{infodomdefn}), then  for  any setwise causal arc-signal vector $S$ (Def. \ref{causaldefn})
and positive integer  $m\in\NN$ that
satisfy (\ref{Xi}) and (\ref{demand}), 
$\exists$ a function $\gamma$ such that  
\beq
S_\alpha(0:km)= \gamma\left (k, S_\mbB(0:km)\right ), \ \ \forall k\in\NN.
\label{harveythmeq}
\eeq

Conversely, 
if for any setwise causal  $S$ 
that meets  (\ref{Xi}) -- (\ref{demand}) with  block-length $m$
there is  a function $\gamma$ ensuring that  (\ref{harveythmeq}) holds,
 then  $\alpha\in\Dom(\mbB)$.
 
\end{theorem}

{\em Remarks:}  The property specified in (\ref{harveythmeq}) is a version of the  concept of {\em informational dominance} 
introduced in \cite{}; the important of this result lies in giving this functional concept a purely structural characterisation. 
The proof follows similar lines as that of Theorem 10 {in} \cite{harveyTIT06}
and is omitted. Minor differences are that $m$ is not constrained to be 1 here, 
and that cyclic networks are handled  using the notion of setwise causality (Def. \ref{causaldefn}),
rather than by introducing channel delays and  then `unwrapping' the network over time to yield
an infinite directed acyclic graph.  

\section{Main Result}
\label{mainsec}

The main result of this paper is presented in this section.
In order to do so, several nonstandard graph-theoretic notions are needed.
Throughout this section, 
$\Sigma=(\mbV,\mbA,\mbP)\equiv (\mbV,\mbAf,\mbAi,\mbP)$ is the structure of an $n$-pairs information network
as described in  Sect. \ref{forsec}, 
and $\Gamma=(\Sigma,S)$ is its setwise-causal signal graph (Def. \ref{causaldefn}),
with source- and sink-signal vectors $X$ and $Y$. 

\begin{definition}[$\mbJ$-Disjointness]
\label{Jdisjointdefn}
Given an index set $\mbJ\subseteq[1:n]$,  an arc set $\mbB\subseteq\mbA$ 
is  $\mbJ${\em-disjoint}  if each path in the $\mbJ$-bundle 
passes through at most one arc in $\mbB$.

If $\mbJ=\{i\}$ for some $i\in[1:n]$, then $\mbB$ 
is called $i${\em-disjoint}.
%%The family of  $\mbJ$-disjoint sets of finite-capacity arcs $\in\mbAf$ is denoted $\mcA_\mbJ$. 
{\flushright $\Diamond$} 
\end{definition} 
{\bf Remarks:} It is easy to see that empty and  singleton arc-sets are automatically $\mbJ$-disjoint,
that every $\mbB\subseteq\mbA$ is  $\emptyset$-disjoint, and that every subset 
of a $\mbJ$-disjoint set inherits its  $\mbJ$-disjointness.
With a little effort, it can also be shown that  $\mbJ$-disjoint arc-sets satisfy the `augmentation' property.
Thus  $\mbJ$-disjoint sets form a {\em finite matroid} 
on $\mbA$.   

Structural dominance (Def. \ref{infodomdefn}) and $[1:i]$-disjointness 
are next used to define  nested families of arc-sets with certain structural properties.
These properties are needed later to inductively extract
{\em entropically feasible} multicommodity flows (Def. \ref{inffeasdefn}).
First, for any arc-set $\mbE\subseteq\mbA$ and $h\in [1:n]$ define  
the  source-augmented set 
\beq
\mbE_{*h}:=\mbE\cup\OUT\lp\sigma_{\mbJ^{h-1}\cup [h+1:n]}\rp,
 \  \ \mbJ^{h-1}\equiv \left\{ j\in [1:h-1]: \mbE\cap\mbA^j = \emptyset\right\}.
\label{defEstar}
\eeq
That is, $\mbE$ is augmented by those source-arcs that either have indices greater than $h$ or 
 that have indices less than $h$ but no  source-sink paths going through $\mbE$.  

\begin{definition}[$i$-Downward Dominated Sets]
\label{idownmatchdefn}
For each $i\in[1:n]$, the family $\mcD_i$ consists of all
arc sets $\mbE\subseteq\mbA$ such that
\begin{enumerate}  
\item $\mbE$ is $[1:i]$-disjoint (Def. \ref{Jdisjointdefn}), and
\item  for each $h\in [1:i]$, either the $h$-bundle does not touch $\mbE$, i.e.
$\mbE\cap\mbA^{h}=\emptyset$, or else the source-augmented arc-set $\mbE_{*h}$ 
(\ref{defEstar}) structurally dominates  the source-arc $\out(\sigma_h)$  (Def. \ref{infodomdefn}).
\end{enumerate}
Every member-set of $\mcD_i$ is called  
  $i${\em-downward dominated}.
{\flushright $\Diamond$} 
\end{definition}

{\bf Remark:} Clearly, every $\mcD_i$-set is also in $\mcD_{i-1}$.

The next concept describes a class of $i$-cuts
that have a special structure: 
\begin{definition}[Viable $i$-Cuts] 
\label{viabledefn} 
Given an index $i\in[1:n]$,
   an $i$-cut $\mbU\subset\mbV^i$  
is called  {\em viable}  
under the following conditions:
\begin{enumerate}
\item Every  arc leaving $\mbU$ in the $i$-bundle is finite-capacity, i.e.  
$\OUT(\mbU)\cap\mbA^i\subseteq\mbAf$.
\item There is an $i$-path that leaves $\mbU$ without re-entering.
\item   Each arc in $\OUT(\mbU)\cap\mbA^i$ lies    
in an $i$-path that either exits $\mbU$ without re-entering or else lies in the $[1:i-1]$-bundle.
\item Every vertex $\nu\in\mbU$ lies on an undirected path $\pi$ from $\sigma_i$ to $\nu$ such that 
\begin{enumerate}
\item all vertices before $\nu$ on $\pi$ are in $\mbU$, and 
\item every reverse-oriented arc in $\pi$ (i.e. pointing from $\nu$ to $\sigma_i$) lies on an $i$-path that does not re-enter $\mbU$.
\end{enumerate}
%%Simplifiy this to only consider vertices $\nu$ in the out-boundary of $\mbU$? 
%%Argument: consider any vertex $\mu$ in $\
\end{enumerate}
{\flushright $\Diamond$}  
\end{definition}

{\bf Remark:} Viable $i$-cuts correspond to  possible {\em min-cuts} in a residual capacitated digraph that 
is used to prove the main result of this chapter (Thm. \ref{mainthm}).
Further investigation of these min-cuts 
may  yield other structural properties to add to the list above;
however, this is left for future work.

%%\begin{definition}[Reverse Structure]
%%\label{revdefn}
%% Given a structure
%%$\Sigma$,
%%the {\em reverse structure}  
%% $\Sigma':=(\mbV,\mbAf',\mbAi',\mbP')$ is given by 
%%\begin{align}
%%\mbAf' &:= \left\{(\mu',\nu'): (\nu',\mu')\in\mbAf\right\},  
%%\label{defmbAfp}\\
%%\mbAi' &:= \left\{(\mu',\nu'): (\nu',\mu')\in\mbAi\right\},  
%%\label{defmbAip}\\
%%\mbP' &:= \left\{(\sigma_i',\tau_i'): (\tau_i',\sigma_i')\in\mbP\right\}.  
%%\label{defmbPp}
%%\end{align}
%%{\flushright $\Diamond$} 
%%\end{definition}
%%{\bf  Remarks:}
%%This describes the $n$-pairs information network obtained by reversing arcs and swapping the roles of sources and sinks.
%%The reverse structure is useful because a multicommodity flow   
%%on $\Sigma$ can be reversed in direction to yield one on $\Sigma'$,  and vice-versa.
%%Markov-like?
%%Markov choke?
%%Spatially Pseudo-markov? 
%%Staggered?
\begin{definition}[Downward Dominance]
\label{tridefn}
%%The capacitated input-output graph $(\mbV,\mbA,\mbS,\mbN,\mbT,c)$ for 
A structure
 $\Sigma$ 
 is called {\em downward dominated} 
if  for each $i\in[2:n]$ and  viable $i$-cut $\mbU$ (Def. \ref{viabledefn}), 
the set $\mbO^i=\OUT(\mbU)\cap\mbA^i$ of outgoing arcs in the $i$-bundle satisfies
the following two conditions:  
\begin{enumerate}
\item $\mbO^i\in\mcD_{i-1}$
%%contains a (possibly empty) 
%%arc set 
(Def. \ref{idownmatchdefn}),
%% $\mbB\supseteq\mbO^i\cap\mbA^{[1:i-1]}$, 
and 
\item the source-augmented arc-set $\mbO^i_{*i}$ (\ref{defEstar}) structurally dominates 
the source-arc $\out(\sigma_i)$  (Def. \ref{infodomdefn}).
\end{enumerate}
{\flushright $\Diamond$}  
\end{definition}
{\bf Remarks:} Note that $1$-pair structures are automatically downward  dominated,
since the conditions above become empty. 
%%ownward dominance is not generally invariant under reversals of structure.
%%Although a viable $i$-cut $\mbU$ corresponds exactly to  
%%a viable  $i$-cut $\mbUc$ in the reverse structure $\Sigma'$,
%%a downward whole $\{\sigma_1,\ldots ,\sigma_{i}\}\leadsto\tau_i$-choke becomes 
%%a $\sigma_i'\leadsto \{\tau_1',\ldots ,\tau_{i}'\}$-choke,
%%{\em not} a  $\{\sigma_1',\ldots ,\sigma_{i}'\}\leadsto\tau_i'$-choke.
 
%%Or, every irreducible fc $i$choke {\em contains}  a $[1:i-1]$-disjoint, 
%%$\{\sigma_1,\ldots ,\sigma_{i-1}\}\leadsto\tau_i$-choke?  
%%No - i need the subset to also be an $i$choke.

A sequence of simpler and progressively more restrictive sufficient conditions for downward dominance can be found
by exploiting Lemma \ref{sdomlem}: 

\begin{lemma}[Simpler Condition 1] 
\label{simplelem}
Suppose that for each $i\in[2:n]$
and  viable $i$-cut  $\mbU\subset\mbV^i$  (Def. \ref{viabledefn}),
the arc-set $\mbO^i=\OUT(\mbU)\cap\mbA^i\subseteq\mbAf$ satisfies the following conditions: 
\begin{enumerate}
\item  $\mbO^i$ is $[1:i-1]$-disjoint (Def. \ref{Jdisjointdefn}), and 
\item for each index $h\in [1:i]$ for which there is a $h$-path that passes through $\mbO^i$, 
i.e. $\mbO^i\cap\mbA^{h}\neq\emptyset$,
all indirect $h$-walks (Def. \ref{indirectdefn})  
pass through the source-augmented arc-set $\mbO^i_{*h}$ (\ref{defEstar}).
\end{enumerate}
Then $\Sigma$ is downward dominated (Def. \ref{tridefn}).
\end{lemma}

\begin{proof}
Follows immediately from applying Lemma \ref{sdomlem} to
Defs. \ref{idownmatchdefn} and \ref{infodomdefn}.
\qed
\end{proof}

\begin{lemma}[Simpler Condition 2]
\label{simplerlem}
Suppose that for each $i\in[2:n]$
and  viable $i$-cut  $\mbU\subset\mbV^i$  (Def. \ref{viabledefn}),
the arc-set $\mbO^i=\OUT(\mbU)\cap\mbA^i\subseteq\mbAf$ satisfies the following conditions: 
\begin{enumerate}
\item  $\mbO^i$ is $[1:i-1]$-disjoint (Def. \ref{Jdisjointdefn}). 
\item For every $h\in [1:i]$ and $s\in[1:h]$ such that 
%%$\mbO^i\cap\mbA^{h}\neq\emptyset$,
%%all $h$-paths go through $\mbO^i$. 
%% and $\mbO^i\subseteq\mbA^h$.
  $\mbO^i\cap\mbA^h$ and $\mbO^i\cap\mbA^s\neq\emptyset$,
%%reachable from $\sigma_s$
 all paths from $\sigma_s$  to $\tau_h$  pass through $\mbO^i$.
%%or none of them do.
\end{enumerate}
Then $\Sigma$ is downward dominated (Def. \ref{tridefn}).
\end{lemma}
\begin{proof}
Let $\mbO^i\cap\mbA^{h}\neq\emptyset$ for some $h\in[1:i]$.
It is asserted that  
all indirect $h$-walks   (Def. \ref{indirectdefn})  must pass through $\mbO^i_{*h}$.

To see this, suppose in contradiction that there is an indirect $h$-walk $\ome$  
that does not pass through  
$\mbO^i_{*h} \equiv  \mbO^i\cup\OUT \lp\sigma_{\mbJ^{h-1}\cup [h+1:n]}\rp$,
where  $\mbJ^{h-1}\equiv \left\{ j\in [1:h-1]:\right.$ 
$\left. \mbO^{i}\cap\mbA^j = \emptyset\right\}$.
Let $\sigma_s$ be the last source vertex in $\ome$,
and let $\pi$ be the subpath from $\sigma_s$ to $\tau_h$. 
Clearly, $s\notin\mbJ^{h-1}\cup [h+1:n]$.
In addition, $s\neq h$, since otherwise $\ome$ reduces to a path from $\sigma_h$ to $\tau_h$, 
which by the second condition above  must pass through $\mbO^i\subseteq\mbO^i_{*h}$.

Thus $s\in [1:h-1]\setminus\mbJ^{h-1}$,  
i.e. $\mbO^i\cap\mbA^s\neq\emptyset$.
%%Consequently, there must be an arc $\alpha\in\mbO^i$ that lies in 
%%an $s$-path. 
%%Thus $\alpha$ is reachable from $\sigma_s$. 
%%Let $\pi_1$ be its subpath from $\sigma_s$ to the tail of $\alpha$.
%%By the second condition above,  
%%As $\alpha$ must then also lie in a $h$-path,
%%Let $\pi_2$ be its subpath from the tail of $\alpha$ to $\tau_h$.
%%Then the concatenated path $\pi_1\pi_2$ goes from $\sigma_s$ to $\tau_h$.   
By the second condition above,  all $\sigma_s\leadsto\tau_h$-paths  must then pass through 
 $\mbO^i$. As $\pi$ is such a path, the indirect $h$-walk $\ome$, of which it is a part, 
passes through $ \mbO^i\subseteq\mbO^i_{*h}$, yielding a contradiction.

The result then follows from Lemma \ref{simplelem}.
\qed
\end{proof} 

\begin{lemma}[Simpler Condition 3]
\label{simplestlem}
Suppose that for each $i\in[2:n]$
and every  viable $i$-cut  $\mbU\subset\mbV^i$  (Def. \ref{viabledefn}),
 there is exactly arc in $\mbO^i=\OUT(\mbU)\cap\mbA^i\subseteq\mbAf$. 

Furthermore, suppose that 
for each $h\in [1:i]$ and $s\in[1:h]$ such that 
%%$\mbO^i\cap\mbA^{h}\neq\emptyset$,
%%all $h$-paths go through $\mbO^i$. 
%% and $\mbO^i\subseteq\mbA^h$.
 $\mbO^i\cap\mbA^h$ and $\mbO^i\cap\mbA^s\neq\emptyset$,
%%reachable from $\sigma_s$
all paths from $\sigma_s$  to $\tau_h$ pass through $\mbO_i$, or none of them do.

Then $\Sigma$ is downward dominated (Def. \ref{tridefn}).
\end{lemma}
\begin{proof}
Observe that $\mbO^i$ consists of a single arc $\alpha$. 
Thus the first condition of Lemma \ref{simplerlem} is trivially satisfied.
To show that its second condition is also met, 
suppose that $\mbO^i\cap\mbA^{h},  \mbO^i\cap\mbA^s\neq\emptyset$ for some $h\in[1:i]$, $s\in[1:h]$.
Thus $\alpha$ is on both an $s$-path and a $h$-path. Let $\pi_1$ be the subpath of the $s$-path from
$\sigma_s$ to the tail of $\alpha$ and $\pi_2$,  the subpath of the $h$-path from
the head of $\alpha$ to $\tau_h$. Then the concatenation $\pi_1\alpha\pi_2$
is a $\sigma_s\leadsto\tau_h$-path that passes through $\mbO^i$.    
By the all-or-nothing condition above, all 
$\sigma_s\leadsto\tau_h$-paths then pass through $\mbO^i$.  
The result then follows from Lemma \ref{simplerlem}.
\qed
\end{proof}

The main result of this chapter can now be stated: 
\begin{theorem}[Downward Dominance $\Rightarrow$ Structural Routability]
\label{mainthm}
If there is an ordering of the source-sink pairs in an $n$-pairs network
 so that the structure  $\Sigma$ is  downward dominated (Def. \ref{tridefn}),
then the achievability of $(X,c)$  (Def. \ref{achdefn}) implies the
existence of  
an $(X,c)$-feasible multicommodity flow  (\ref{mccapbnd})--(\ref{mccons}).

Conversely, if  $X$ is stationary and there exists an $(X,c)$-feasible multicommodity flow  
with (\ref{mccapbnd}) holding in strict form, then
$(\Sigma, X,c)$ is achievable.
$\bigtriangledown$ 
\end{theorem}

%%{\bf Remarks:} The primary  significance of this result is structural, not computational.
%%Algorithms can be trivially devised to check whether or not a given network topology is downward dominated,
%%by working directly from the definition. However, the explicit construction of such a procedure, and the analysis of its computational
%%complexity, is not addressed here.

%%In addition, although downward dominance  is sufficient to guarantee 
%%that routing can always achieve any feasible combination of source signals, demand rates and channel capacities on a given topology,
%%it may not be  necessary. If it is not necessary, then the question remains 
%%of finding a  topological condition that is less conservative, or even tight.
%%This is also left for future work. 

{\bf Remarks:}
This result  
defines a non-trivial class of directed network structures
 for which achievability is essentially equivalent to the existence of a  feasible multicommodity flow.
On these structures, information can indeed be treated 
like an incompressible, immiscible fluid flow.
  
The proof of Thm. \ref{mainthm} is given in 
the next section.
In Sect. \ref{exsec}, several network examples are discussed to illustrate the  applicability of Thm. \ref{mainthm}.

\section{Proof of Theorem \ref{mainthm}}
\label{keysec} 

%% and
%%$\Gamma=(\Sigma,S)$, a  signal graph, 
%%and Def. \ref{causaldefn}.
In both the proofs of necessity and sufficiency,
 use will be made of the fact that  $\forall i\in[1:n]$, any single-commodity flow $q$ from $\sigma_i$ to $\tau_i$
in the structure $\Sigma$ 
 can be decomposed into a superposition
of $i${\em-path flows} and {\em cycle flows} (see e.g. \cite{bang}, Thm. 3.3.1).  
That is, if $\pi_{1,i},\ldots, \pi_{p_i,i}$ are the distinct $i$-paths   
and $\gamma_1,\ldots, \gamma_g$, the distinct cycles,
then  $\exists$  numbers  $u_{1,i},\ldots , u_{p,i}\geq 0$ and $w_{1,i},\ldots , w_{g,i}\geq 0$ s.t.
\beq
q_{\alpha} =   
\sum_{1\leq k\leq p_i: \pi_{k,i}\ni\alpha } u_{k,i} +  \sum_{1\leq l\leq g: \gamma_l\ni\alpha } w_{l,i}. 
\label{decomp}
\eeq
If $w_{l,i}=0$ for all $l\in[1:g]$, then  the flow $q$ is called  {\em acyclic}.

The proof of sufficiency in Sect. \ref{suffsubsec} is relatively straightforward.
Given an $(X,c)$-feasible multicommodity flow $f$ (\ref{mccapbnd})--(\ref{mccons}) on $\Sigma$, 
the decomposition (\ref{decomp}) is used directly to devise a routing solution $S$. 

The proof of necessity in Sect. \ref{necsubsec} is more difficult and involves induction,
using the following building blocks. 
%%The following notion is a basic building block in the proof is the following notion:
\begin{definition}[$\mbJ$-Flow] 
\label{Jflowdefn}
Given an index set $\mbJ\subseteq[1:n]$, a nonnegative tuple $f=\left (f_{\alpha,j}\right )_{\alpha\in\mbA, j\in\mbJ}\in\RR^{|\mbA||\mbJ|}_{\geq 0}$ 
is called a $\mbJ${\em-flow} on the structure $\Sigma$  
if 
 $\forall j\in\mbJ$ and  $\nu\in\mbV\setminus\{\sigma_j\}\cup\{\tau_j\}$, 
\beq  
\sum_{\alpha\in\IN(\nu)}f_{\alpha,j} 
=   \sum_{\alpha\in\OUT(\nu) }f_{\alpha,j} 
 \ \ \mbox{($j$-flow conservation)}, 
\label{Jcons}
\eeq
As a convention, the $\emptyset$-flow is defined as the empty sequence $()$. {\flushright $\Diamond$} 
\end{definition}
{\bf Remark:} A $\mbJ$-flow is  a (possibly infeasible) multicommmodity flow   
with source-sink pairs $(\sigma_j, \tau_j)$, $j\in\mbJ$.
%%If $w_{l,i}=0$ for all $l\in[1:g]$, then  $q$ is called an {\em acyclic $j$-flow},
If each $j$-flow $f_{\mbA,j}$ is acyclic, $\forall j\in\mbJ$, then $f$ is called an {\em acyclic $\mbJ$-flow}.

%%\begin{lemma}
%%\label{subflowlem}
%%If $\mbK\subseteq\mbJ\subseteq\{1,\ldots ,n\}$,
%%then any $\mbJ$-flow is also an $\mbK$-flow.
%%\end{lemma}
%%
%%{\em Proof:}   Trivial. 
%%$\Box$

%%\begin{lemma}
%%\label{setflowlem}
%%For any $\mbJ$-flow and any internal vertex set $\mbU\subseteq\mbVo$,
%%\[
%%\sum_{\alpha\in\IN(\mbU)}f_{\alpha,j} =   \sum_{\alpha\in\OUT(\mbU) }f_{\alpha,j}, 
%% \ \forall \mbU\subseteq\mbVo, j\in\mbJ.
%%\ \ \bigtriangledown
%%\]
%%\end{lemma}

%%{\em Proof:} 
%%Trivial. 

%%Wrong!  
%%\begin{lemma}
%%Every minimal $\mbJ$-choke is $\mbJ$-disjoint.
%%\end{lemma}

%%{\em Proof:} Suppose in contradiction that $\exists$ a minimal $\mbJ$-choke $\mbB$ that is not $\mbJ$-disjoint.
%%Then $\exists$ two arcs $\alpha_1,\alpha_2\in\mbB$  that lie on the same $h$-path, with $h\in\mbJ$.  
The next concept is central to the proof of necessity.
It defines a class of feasible $[1:i]$-flows 
that obey  certain information-theoretic bounds
when 
%%$X_{[i+1:n]}$ is made globally available,  
only the signals $X_j$, $j\in[1:i]$,  need to be communicated.
\begin{definition}[Entropic Feasibility] 
\label{inffeasdefn}
%%Consider an $n$-pairs information network
%%with signal graph $\Gamma=(\mbV,\mbAf,\mbAi,\mbP,\mbN,S)$.
Given  $i\in[1:n]$ and a solution $S$ to $(\Sigma,X,c)$ (Def. \ref{achdefn}), 
a $[1:i]$-flow $f\in\RR^{|\mbA| i}_{\geq 0}$  (Def. \ref{Jflowdefn})  
is called {\em entropically feasible}    
if it satisfies the following conditions: 
\begin{itemize}
\item[i)] On every arc  $\alpha\in\mbAf$,
\beq
\sum_{j=1}^i f_{j,\alpha}\leq c_\alpha. 
\label{capbndinf}
\eeq
\item[ii)] On any  $i$-downward dominated arc set $\mbB$ (Def. \ref{idownmatchdefn}), 
\beq
\sum_{\alpha\in\mbB,j\in[1:i]} f_{\alpha,j}\leq  \Hsup\left [S_\mbB | X_{\mbJ^{i}\cup[i+1:n]}\right ],
\label{infbnd}
\eeq
where $\mbJ^{i}= \left\{ j\in [1:i]: \mbB\cap\mbA^j = \emptyset\right\}$. 
\item[iii)] On  arcs entering sinks and leaving sources,  
\beq 
f_{\into(\tau_j),j} = f_{\out(\sigma_j),j}  =   
%% \left\{\begin{array}{ll}
   \Hinf [X_j], 
\  \  \forall j\in[1:i].
\label{flowdemand} 
\eeq
\end{itemize}
{\flushright $\Diamond$} 
\end{definition}

%%By convention, these conditions are deemed satisfied by the empty tuple when $\mbJ=\null$.
{\bf Remarks:}  
%%Intuitively, (\ref{infbnd}) is restricted to $[1:i]$-disjoint sets  because the flow contribution
%%corresponding to a path in the $[1:i]$-bundle could otherwise be counted more than once.
Note that the $\emptyset$-flow is entropically feasible,
since the condition (\ref{flowdemand})  disappears and
 (\ref{infbnd}) is trivially satisfied due to a zero left-hand side (LHS).

The proof of necessity in the next section  proceeds by
constructing 
 an entropically feasible  $[1:n]$-flow    
on $(\Sigma,c,S)$, which automatically  
gives the desired  
 $(X,c)$-feasible multicommodity flow
(\ref{mccapbnd})--(\ref{mccons}).

%%The proof of necessity  uses upward induction to construct
%%a sequence of entropically feasible $[1:i]$-flows (Def. \ref{Jflowdefn}), $i\in[1:n]$.
%% By Lemma \ref{1nflowlem}, $f^n$ is then an $(X,c)$-feasible multicommodity flow,
%%as desired.

\subsection{Necessity Proof for Theorem \ref{mainthm}}
\label{necsubsec} 

Let the arc-signal vector $S$ be a solution (Def. \ref{achdefn}) to
the $n$-pairs information network problem $(\Sigma,X,c)$. 
An entropically feasible $[1:n]$-flow (Def. \ref{inffeasdefn}) $f^n$ will  be constructed, using 
upward induction. 

Let $\Sigma$ be downward dominated (Def. \ref{tridefn})
and suppose that  
 $f^{i-1}=\left (f_{\alpha,j} \right )_{\alpha\in\mbA,j\in[1:i-1]}$ $\in\RR^{|\mbA| (i-1)}_{\geq 0}$ 
is an entropically feasible, acyclic $[1:i-1]$-flow
for some $i\in [1:n]$, 
noting that the  $\emptyset$-flow $f^0$ is entropically feasible.
%% nonnegative vector $\in\RR^{|\mbA|n}_{\geq 0}$
%%is an entropically feasible $\emptyset$-flow.
%% set $f_{\alpha,j}^i=f_{\alpha,j}$, $\forall \alpha\in\mbA, j\in[1:i-1]$.
An $i$-flow $\left (f_{\alpha,i}\right )_{\alpha\in\mbA}\in \RR^{|\mbA|}_{\geq 0}$ 
will be constructed in such a way that  
$f^i \in\RR^{|\mbA| i}_{\geq 0}$ will be an entropically feasible, acyclic $[1:i]$-flow.

On any arc $\alpha\in\mbA$, let  
\beq 
r_\alpha:= \left\{\begin{array}{ll}
c_\alpha - \sum_{j=1}^{i-1} f_{\alpha,j}  
&\mbox{if }\alpha\in\mbAf\\
\infty &\mbox{if }\alpha\in\mbAi\equiv\mbA\setminus\mbAf 
\end{array}\right.
\label{defralpha}
\eeq
be the residual capacity after subtracting the relevant components of $f^{i-1}$.  
%% and 
%%let 
%%\[
%%m_{i-1}(\mbB):=  \Iinf [X_{[1:i-1]}; S_\mbB | X_{[i:n]},W] - \sum_{j\in[1:i-1],\alpha\in\mbB}f_{j,\alpha},
%% \ \ \forall \mbB\in\mcA_{[1:i-1]}.
%%\]
Note that  $r_\alpha\stackrel{(\ref{capbndinf})}{\geq}0$ since $f^{i-1}$ is an entropically feasible $[1:i-1]$-flow.  
%%From Lemma \ref{pseudentlem},  $m_{i-1}$ is a pseudo-entropy on $\mcA_{[1:i-1]}$.
%%Thus 
%%\[
%%\sum_{j=1}^{i-1} f_{\alpha,j} \leq  
%% c_\alpha.
%%\]
%%Thus $r_\alpha\geq 0$, $\forall\alpha\in\mbA$.
The next step is to find an acyclic $i$-flow (Def. \ref{Jflowdefn}) $q\in\RR^{|\mbA|}_{\geq 0}$ from $\sigma_i\leadsto\tau_i$ 
that is {\em a)} $\leq$ the residual capacity on each arc,  and {\em b)}
 $\geq \Hinf [X_{i} ]$  on the arc entering $\tau_i$.
%%where 
There are two mutually exclusive cases to consider.

\subsubsection{1st Case: $\exists$ an $i$-Path with No Finite-Capacity Arcs}
\label{1stcasepar}
%%There is an $i$-path with all arcs $\in\mbAi$, i.e. all its arcs have  infinite capacity.
Denote this  $i$-path by $\pi_e$,  noting that $r_\alpha=\infty$, $\forall\alpha\in\pi_e$ by the 2nd line of (\ref{defralpha}).
Set the $i$-path flows as  
\beq
 u_k=\left\{\begin{array}{ll}
 \Hinf [X_{i}]
 &\mbox{if } k=e\\
0 &\mbox{otherwise}
\end{array}\right., \  \forall k\in [1:p],  
\label{udegen}
\eeq
and the cycle flows equal to zero 
in the decomposition (\ref{decomp})
(dropping the $i$-subscripts),  
so that 
\beq
q_\alpha \stackrel{(\ref{decomp})}{=}   
\sum_{1\leq k\leq p: \pi_k\ni\alpha } u_k,
\ \ \forall\alpha\in\mbA. \label{decomp3}
\eeq
Evidently $q$ is acyclic and meets the residual capacity constraint 
on all arcs in $\mbA$.
Furthermore, since every $i$-path passes through the single arc entering $\tau_i$,
\beq
q_{\into(\tau_i)}  \stackrel{(\ref{decomp3})}{=} \sum_{1\leq k\leq p } u_k
  \stackrel{(\ref{udegen})}{=}  u_e =  \Hinf [X_{i} ],
\label{qI1}
\eeq
satisfying the conditional information constraint.
 
\subsubsection{2nd Case: Every $i$-Path Has One or More Finite-Capacity Arcs}
\label{2ndcasepar}
%%Every $i$-path has at least one finite-capacity arc $\in\mbAf$.
Observe first that for any arc set $\mbB\subseteq\mbA$,
\begin{align}
\sum_{\beta\in\mbB} c_\beta 
&\stackrel{(\ref{capbnd})}{\geq}   \sum_{\beta\in\mbB } 
\Hsup[S_\beta] 
 \equiv \sum_{\beta\in\mbB } \varlimsup_{t\to\infty}\frac{\mrH\left [S_\beta(0:t) \right ]}{t+1}
\nn\\
&\geq   \varlimsup_{t\to\infty}\frac{1}{t+1}\sum_{\beta\in\mbB }  \mrH\left [S_\beta(0:t) \right ]
 \geq   \varlimsup_{t\to\infty}\frac{\mrH\left [S_\mbB(0:t) \right ] }{t+1}  
\equiv \Hsup [S_\mbB] 
\label{Hsumbnd}\\ 
 &\geq   \Hsup \left [S_\mbB | X_{\mbJ^{i-1}\cup [i+1:n]}\right ]   
\label{Hcondbnd}\\
&=   \varlimsup_{t\to\infty}
\left (\frac{\mrH\left [S_\mbB(0:t)| X_{\mbJ^{i-1}\cup[i+1:n]}(0:t) \right ]
- \mrH\left [S_\mbB(0:t)| X_{\mbJ^{i-1}\cup[i:n]}(0:t) \right ]}{t+1}\right.\nn\\
&  \left.\mbox{}  + \frac{\mrH\left [S_\mbB(0:t)| X_{\mbJ^{i-1}\cup[i:n]}(0:t) \right ]}{t+1}
\right )   
\nn\\
&=  \varlimsup_{t\to\infty}\left (\frac{\mrI\left [S_\mbB(0:t); X_{i}(0:t)| X_{\mbJ^{i-1}\cup[i+1:n]}(0:t) \right ]}{t+1}\right.\nn\\
&  
\left. \mbox{} + \frac{\mrH\left [S_\mbB(0:t)| X_{\mbJ^{i-1}\cup[i:n]}(0:t) \right ]}{t+1}\right ) 
\nn\\
&\geq   \Iinf\left [X_{i}; S_\mbB | X_{\mbJ^{i-1}\cup[i+1:n]} \right ] 
  + \Hsup\left [ S_\mbB | X_{\mbJ^{i-1}\cup[i:n]} \right ]
\nn\\
&= \Iinf\left [X_{i}; S_\mbB , X_{\mbJ^{i-1}\cup[i+1:n]} \right ] 
  + \Hsup\left [ S_\mbB | X_{\mbJ^{i-1}\cup[i:n]} \right ], 
\label{Isplit}
\end{align}
%%In (\ref{Hdeterm}) and (\ref{Hdeterm2}), the conditional discrete entropies are 0 since, by Lemma \ref{reachlem}, $S_\mbB(0:t)$ 
%%is a function of  $X(0:t)$; 
where  (\ref{Hsumbnd}) is due to the subadditivity of joint entropy,
  (\ref{Hcondbnd}) holds because conditioning cannot increase entropy,
and (\ref{Isplit}) arises from  the mutual independence of $X_1,\ldots , X_n$.

Now, consider the residual capacitated digraph $\left (\mbV^i,\mbA^i,r_{\mbA^i}\right )$
formed by the $i$-bundle.\footnote{Here, arcs are permitted to have $r_\alpha=0$.}   
%%with single source-sink pair $(\sigma_i,\delta_i)$. 
Let $q$ be an acyclic maximal flow on it
under the constraints  
\beq
0\leq q_{\alpha} \leq  r_\alpha, \ \ \forall\alpha\in\mbA^i. \label{qfeas}
\eeq
By the {\em Min-Cut Max-Flow Theorem} (see e.g. \cite{bang}, Thm. 3.5.3)
$\exists$ an
$i$-cut $\mbU\subset\mbV^i$,
consisting of  
every vertex $\nu\in\mbV^i$ for which 
$\exists$ an undirected path $\pi$ in $(\mbV^i,\mbA^i)$ from $\sigma_i$ to $\nu$ 
s.t.   
\begin{itemize}
\item {\em (Forward Slack)} every forward-oriented arc $\alpha$ in $\pi$ (i.e. pointing from $\sigma_i$ to $\nu$)
    has $q_\alpha < r_\alpha$, and 
\item {\em (Backward Flow)}  every backward-oriented arc $\alpha$ in $\pi$ (pointing from $\nu$ to $\sigma_i$) has $q_\alpha>0$.
\end{itemize} 
As a consequence of this, 
%%Wait...the min-cut max-flow thm is fairly powerful.
%%In this step, i am applying it not to the oiginal network, but a modified one.
%%Perhaps I can make a better modification?
%%The triangularity condition i have now is assymmetric between sources and sinks.
%%However, by symmetry, if i reverse the flows and the roles of sources and sinks, i should
%%still get a flow.
%%So i want a condition that is also symmetric! 
%%How about this: do as above with $r_\alpha$, then add zero capacity links from the successor of $\sigma_i$ 
%%to the successors of each $\sigma_j$, $j\in[1:i-1]$  and zero-capacity links from the predecessors of $\tau_j$, $j\in[1:i-1]$,
%%to the predecessor of $\tau_i$. In other words, we are hijacking the $[1:i-1]$ bundle from a structural point of view,
%%but ensuring that no $i$-flow can actually go directly into the $[1:i-1]$ bundle.
%%Then apply min-cut max-flow. All the structural conditions below hold on $\mbU$. 
%%Almost automatically though, $\OUT(\mbU)$  is, or contains, a $[1:i]$-choke!
%%We need only impose $[1:i-1]$ disjointness...  
\begin{align}
q_{\alpha} &=  r_\alpha, \ \ \forall \alpha\in\mbO^i:=\OUT(\mbU)\cap\mbA^i, \label{qsat}\\
q_{\alpha} &=  0, \ \ \forall \alpha\in\mbI^i:=\IN(\mbU)\cap\mbA^i.\label{qzero}
\end{align}  
Note also that since the cyclic flow components $w_1,\ldots , w_j$ in (\ref{decomp})  
are zero,   
\beq
q_\alpha    
= \sum_{1\leq k\leq p:  \pi_k\ni\alpha } u_k,
\ \ \forall\alpha\in\mbA^i. \label{decomp2}
\eeq

The $i$-cut $\mbU$ evidently depends on the residual capacity vector $r$. However,
the following {\em purely structural} statements may be made about it: 
%%Trivially, every $i$-path  $\pi_k$ must transit through $\OUT(\mbU)$.
\begin{enumerate}
\item  Every arc in $\mbO^i$ lies in $\mbAf$, i.e. is finite-capacity. 
Otherwise  $q_\alpha \stackrel{(\ref{qsat})}{=} r_\alpha \stackrel{(\ref{defralpha})}{=}\infty$,
implying by (\ref{decomp2}) that $u_k=\infty$ on some $i$-path $\pi_k$, which is impossible since every $i$-path in this case
travels over at least one finite-capacity arc. 
\item Every arc $\alpha\in\mbO^i$ is in an $i$-path that exits $\mbU$ without re-entering,  or else $\alpha$ is in the $[1:i-1]$-bundle. 
To see this, 
%%suppose $\alpha\in \OUT(\mbU)$ is not on any $i$-path. 
%%As 
%%\beq
%%c_\alpha - \sum_{j=1}^{i-1} f_{\alpha,j} 
%%\stackrel{(\ref{defralpha})}{\equiv} r_\alpha  \stackrel{(\ref{qsat})}{=} q_\alpha   \stackrel{(\ref{decomp2})}{=} 0,
%%\label{fpos}
%%\eeq
%%the positivity of $c_\alpha$ then implies that  $f_{\alpha,j}>0$ for some $j\in[1:i-1]$.
%%As the $j$-flow $(f_{\alpha,j})_{\alpha\in\mbA}$ is acyclic by construction,
%% it follows from (\ref{decomp}) that 
%%$\alpha$ must then lie on a $j$-path. 
%%observe that if an $i$-path $\pi_k$ passes through several arcs in $\mbO^i$, then each such arc
%%must also be in another $i$-path that only passes through $\mbO^i$ once, or else in a path in the $[1:i-1]$-bundle. 
suppose that every $i$-path  $\pi_k$  passing through 
$\alpha$ re-enters $\mbU$.
Evidently,  it must then pass through  
some arc $\beta\in\mbI^i$. By (\ref{qzero})   
 $q_\beta = 0$, implying by virtue of (\ref{decomp2}) and  nonnegativity
that $u_k=0$. 
From 
(\ref{qsat}) and (\ref{decomp2}), 
this implies that 
$r_\alpha=0$. As $c_\alpha>0$, it must then hold that  $f_{\alpha,j}>0$ for some $j\in[1:i-1]$.   
As the $j$-flow $(f_{\alpha,j})_{\alpha\in\mbA}$ is acyclic by construction,
$\alpha$ must then lie on a $j$-path, by (\ref{decomp}). 
\item There must be an $i$-path that leaves $\mbU$ without re-entering.
To see this, suppose in contradiction that every $i$-path re-enters $\mbU$.
By the preceding argument, all $i$-paths must then have associated acyclic flow components $u_k=0$. 
%%implying by (\ref{decomp2}) that the total  $i$-flow $q_{\out(\sigma_i)}=0$.
Pick any $i$-path and let $\nu$ be the last vertex in $\mbU$ that it traverses before
leaving $\mbU$ without further re-entry.
Let $\ome$ denote its  subpath from $\nu\leadsto\tau_i$. 
By the definition of $\mbU$, there is an undirected path $\pi$ from $\sigma_i$ to $\nu$ 
such that all forward-oriented arcs in it are slack and all backward-oriented arcs carry strictly positive $q$-flow.
Note also that  all vertices before $\nu$ in $\pi$ must also lie in $\mbU$, by construction.
From (\ref{decomp2}), any backward arc in $\pi$ would have to carry an $i$-path flow component $u_k>0$, which would be a contradiction.
Consequently, all the arcs in $\pi$ must be forward-oriented, i.e. $\pi$ is a directed path in $\mbU$ from $\sigma_i\leadsto\nu$.
The concatenation of $\pi$ with $\ome$ then yields an $i$-path that leaves $\mbU$ exactly once, a contradiction.  
\item Finally, by construction of $\mbU$, every vertex $\nu$ in it  must lie on an undirected path $\pi$ from $\sigma_i$ to $\nu$ such that 
\begin{enumerate}
\item every vertex before $\nu$ in $\pi$ is also in $\mbU$ (since the subpath from $\sigma_i$ to $\nu$ automatically satisfies the
defining forward-slack and backward-flow properties), and    
\item every reverse-oriented arc in $\pi$  lies on an $i$-path that does not re-enter $\mbU$ (since such arcs must by definition carry positive $q$-flow,
and $i$-paths that re-enter $\mbU$ carry zero $q$-flow).
\end{enumerate}
\end{enumerate}
In other words, $\mbU$ is a {\em viable $i$-cut} (Def. \ref{viabledefn}). 
By downward dominance (Def. \ref{tridefn}),  $\mbO^i\cup\OUT(\sigma_{\mbJ^{i-1}\cup[i+1:n]})$ 
structurally dominates $\out(\sigma_i)$ (Def. \ref{infodomdefn}),
and $\mbO^i$ is a $\mcD_{i-1}$-set (Def. \ref{idownmatchdefn}).
%%$\mbB\supseteq\mbO^i\cap\mbA^{[1:i-1]}$.
%%the $i$-flow 
%%let
%%\beq
%%q_\alpha:= \sum_{1\leq p\leq l:  \alpha \ \mrm{on} \ \pi_p } u_p.
%%\label{defq}
%%\eeq
%%It is straightforward to establish that the nonnegative vector 
%%$q:=\left (q_\alpha\right )_{\alpha \ \mrm{on} \ i\mrm{-bundle}}$ 
%%is a flow on the $i$-bundle subgraph.
Using $i$-flow conservation, 
\begin{align}
q_{\into(\tau_i)}
 &= \sum_{\beta\in\mbO^i} 
q_\beta   -\sum_{\alpha\in\mbI^i} 
q_\alpha  \stackrel{(\ref{qsat}), (\ref{qzero})}{=}   
\sum_{\beta\in\mbO^i} r_\beta\nn\\
&\stackrel{(\ref{defralpha})}{=} 
\sum_{\beta\in\mbO^i} c_\beta -  \sum_{\beta\in\mbO^i,j\in[1:i-1]} f_{\beta,j} 
\nn\\
%%&=
%%\sum_{\beta\in\mbO^i} c_\beta -  \sum_{\beta\in\mbB,j\in[1:i-1]} f_{\beta,j} 
%%\label{reduc}\\
&\stackrel{(\ref{Isplit})}{\geq}  
\Iinf\left [X_{i}; S_{\mbO^i} , X_{\mbJ^{i-1}\cup[i+1:n]} \right ] 
  + \Hsup\left [ S_{\mbO^i} | X_{\mbJ^{i-1}\cup[i:n]} \right ] 
-  \sum_{\beta\in\mbB, j\in[1:i-1]}f_{j,\beta}
\nn\\ 
%%&\geq   
%%\Iinf\left [X_{i}; S_{\mbO^i} , X_{[i+1:n]} \right ] 
%%  + \Hsup\left [ S_{\mbB} | X_{[i:n]} \right ] 
%%-  \sum_{\beta\in\mbB, j\in[1:i-1]}f_{j,\beta}
%%%\label{reduc2}\\ 
&\stackrel{(\ref{infbnd})}{\geq}  
 \Iinf\left [X_{i} ;S_{\mbO^i} , X_{\mbJ^{i-1}\cup[i+1:n]} \right ].
\label{qinlbnd0}
\end{align}
%%where (\ref{reduc})
%% holds because the acyclic $[1:i-1]$-flow $f^{i-1}$ 
%%has zero-valued components on all arcs not in $\mbA^{[1:i-1]}$.
%%and the bound (\ref{reduc2}) follows from the monotonicity of entropy.
As  $\out(\sigma_i)\in\Dom\lp \mbO^i\cup\OUT(\sigma_{\mbJ^{i-1}\cup[i+1:n]})\rp$,
it follows that 
$X_i(0:km-1)$ 
%%$\lp =Y_i(0:km-1)\rp $ 
is a function of 
$S_{\mbO^i}(0:km-1)$ and $X_{\mbJ^{i-1}\cup[i+1:n]}(0:km-1)$.  
Consequently, $\forall k\in\NN$,   
\[
\mrI\left [X_{i}(0:km-1) ;S_{\mbO^i}(0:km-1),  X_{\mbJ^{i-1}\cup[i+1:n]}(0:km-1) \right ]
=   \mrH\left [X_{i}(0:km-1) \right ]. 
\]
As entropy and mutual information are monotonic, 
a sandwich argument with $k\to\infty$ then yields that the RHS of (\ref{qinlbnd0}) is just $\Hinf[X_i]$,
so that 
\beq
q_{\into(\tau_i)}\geq  \Iinf\left [X_{i} ;S_{\mbO^i},  X_{\mbJ^{i-1}\cup[i+1:n]} \right ]=\Hinf [X_{i} ], 
\label{qinlbnd2}
\eeq
as desired.
%%verifying the conditional information constraint. 

\subsubsection{Construction of $f^i$ in Both Cases}
\label{constrpar}
For both cases above, let   
\beq
f_{\alpha,i} :=  \underbrace{\frac{\Hinf [X_{i} ]}{q_{\into(\tau_i)}}}_{=:v}q_\alpha \equiv vq_\alpha,
\ \ \forall \alpha\in\mbA, 
\label{deffi}  
\eeq
where $v\in (0,1]$ by (\ref{qinlbnd2}). 
Clearly, $f_{\mbA,i}$ is still 
an acyclic $i$-flow since it just a scaled version of $q$. 
Furthermore, 
\[
\sum_{j=1}^i f_{\alpha,j} \stackrel{(\ref{deffi})}{=} vq_\alpha + \sum_{j=1}^{i-1} f_{\alpha,j} 
 \stackrel{(\ref{qinlbnd2})}{\leq}   q_\alpha + \sum_{j=1}^{i-1} f_{\alpha,j} \stackrel{(\ref{defralpha})}{\leq}   c_\alpha. 
\]

The next step is to  verify that  $f^i=f_{\mbA\times [1:i]}$
satisfies the remaining conditions (\ref{infbnd})--(\ref{flowdemand}) for  an entropically feasible $[1:i]$-flow.
First (\ref{infbnd}) is checked.
Let $\mbE$ be any arc-set
in $\mcD_{i}$ (Def. \ref{idownmatchdefn}). 
If $\mbE\cap\mbA^{i}=\emptyset$, then   
\begin{align}  
\sum_{\eta\in\mbE, j\in[1:i]}  f_{\eta,j} 
 &=   \sum_{\eta\in\mbE, j\in[1:i-1]}  f_{\eta,j} 
\nn\\
&\stackrel{(\ref{infbnd})}{\leq} \Hsup\left [S_{\mbE} | X_{\mbJ^{i-1}\cup[i:n]}\right ]
 = \Hsup\left [S_{\mbE} | X_{\mbJ^{i}\cup[i+1:n]}\right ]
\nn
\end{align}
since $\mbE\in\mcD_{i-1}$ automatically, and where the last equality  follows
because $ \mbJ^{i-1}\cup \{i\} =\mbJ^{i}$. 
%%&=  \sum_{\eta\in\mbE_1} f_{\eta,i} +  \sum_{\eta\in\mbE_2, j\in[1:i-1]}  f_{\eta,j} 
Else if  $\mbE\cap\mbA^{i}\neq \emptyset$,
write 
%%$\mbE$ must contain sets $\mbE_1,\mbE_2$ and $\mbE_3$ each satisfying the conditions
%%listed in  Def. \ref{idownmatchdefn}.
%%Then 
\beq 
\sum_{\eta\in\mbE, j\in[1:i]}  f_{\eta,j} 
=  \sum_{\eta\in\mbE} f_{\eta,i} +  \sum_{\eta\in\mbE, j\in[1:i-1]}  f_{\eta,j} 
%%&=  \sum_{\eta\in\mbE_1} f_{\eta,i} +  \sum_{\eta\in\mbE_2, j\in[1:i-1]}  f_{\eta,j} 
\label{red1}
\eeq
and bound each sum on the RHS as follows. 
First,  note that since $\mbE\in\mcD_{i-1}$, 
\beq
 \sum_{\eta\in\mbE, j\in[1:i-1]}  f_{\eta,j} 
\stackrel{(\ref{infbnd})}{\leq}   \Hsup\left [S_{\mbE} | X_{\mbJ^{i-1}\cup[i:n]}\right ]. 
\label{secondsum}
\eeq
%%Next, note that if $\mbE\cap\mbA^i=\emptyset$, then 
%%it trivially holds that 
%%\beq
%%\sum_{\eta\in\mbE} f_{\eta,i}
%%=0\leq \Iinf\left [X_{i} ;S_{\mbE} , X_{[i+1:n]} \right ].
%%\label{trivbnd}
%%\eeq
Then write    
\begin{align}
 \sum_{\eta\in\mbE} f_{\eta,i} 
&   \stackrel{(\ref{deffi})}{=}   \sum_{\eta\in\mbE} v q_\eta 
 \stackrel{(\ref{decomp2}), (\ref{decomp3})}{=}   v\sum_{\eta\in\mbE} \left (\sum_{1\leq k\leq p: \pi_k\ni\eta  } u_k\right )  
\nn\\
 &=  v \sum_{1\leq k\leq p }u_k \left (\sum_{\eta\in\mbE:\eta\in\pi_k}1\right )\nn\\
&  \leq    
  v \sum_{1\leq k\leq p } u_k \equiv \nu q_{\into(\tau_i)} \stackrel{(\ref{deffi})}{=}  \Hinf [X_{i} ],
\label{flowbnd1} 
\end{align}
where the inequality  arises  
because the $i$-path flows $u_1,\ldots , u_p\geq 0$ and each $i$-path $\pi_k$ transits over at most one arc in $\mbE$.
As $\out(\sigma_i)\in\Dom\lp \mbE\cup\OUT(\sigma_{\mbJ^{i-1}\cup[i+1:n]})\rp$,
the same arguments that lead to the equality in (\ref{qinlbnd2}) 
show that $\Hinf [X_{i} ] =$ $ \Iinf\left [X_{i} ;S_{\mbE} , X_{\mbJ^{i-1}\cup[i+1:n]} \right ]$. 
Substituting this into (\ref{flowbnd1}) and then combining with (\ref{red1}) and (\ref{secondsum}) yields
\begin{align}
\sum_{\eta\in\mbE, j\in[1:i]}  f_{\eta,j} 
& \leq   \Iinf\left [X_{i} ;S_{\mbE} , X_{\mbJ^{i-1}\cup[i+1:n]} \right ]  +  \Hsup \left [S_{\mbE} | X_{\mbJ^{i-1}\cup[i:n]}\right ]  
\nn\\
& \stackrel{(\ref{Hcondbnd}),(\ref{Isplit})}{\leq}   \Hsup\left [S_\mbE | X_{\mbJ^{i-1}\cup[i+1:n]}\right ]
 = \Hsup\left [S_\mbE | X_{\mbJ^{i}\cup[i+1:n]}\right ],
%%\label{flowbnd2}
\end{align}
since $\mbJ^i=\mbJ^{i-1}$ in this case. 
This 
confirms that $f^i$ satisfies (\ref{infbnd}).
As $f^{i-1}$ is an entropically feasible $[1:i-1]$-flow, 
(\ref{flowdemand}) is satisfied $\forall j\in[1:i-1]$. 
Using flow conservation,   
\[ 
f_{\out(\sigma_i),i} 
= f_{\into(\tau_i),i} 
 \stackrel{(\ref{deffi})}{=} vq_{\into(\tau_i)}= \Hinf [X_{i}  ],
\]
verifying (\ref{flowdemand}) when $j=i$.  Thus  
$f^i$ is an entropically feasible $[1:i]$-flow.
 
By induction,  $f^n$ is  an entropically feasible $[1:n]$-flow,
%%If the reverse structure $\Sigma'$ is downward dominated,
%%the induction goes through identically, but with $(\mbAf',\mbAi',\mbP')$ replacing $(\mbAf,\mbAi,\mbP)$.
%%The  entropically feasible $[1:n]$-flow ${f'}^n$ obtained on $\Sigma'$ then yields an entropically feasible $[1:n]$-flow $f^n$
%%on $\Sigma$, using $f_{(\mu,\nu),j} = f_{(\nu,\mu),j}'$, $\forall (\mu,\nu)\in\mbA$ and $j\in[1:n]$.  
giving the desired  $(X,c)$-feasible multicommodity flow
(\ref{mccapbnd})--(\ref{mccons}).

\subsection{Sufficiency of Multicommodity Flows}
\label{suffsubsec}

The converse part of Thm. \ref{mainthm} is easier to establish,
since it is not difficult to see that the existence of a feasible multicommodity flow
implies  achievability. Thus  only the key steps are provided below.  

Suppose $f$ is an $(X,c)$-feasible multicommodity flow  (\ref{mccapbnd})--(\ref{mccons}) on an $n$-pair network structure
$\Sigma$, with $X$ stationary, and further suppose that (\ref{mccapbnd}) is satisfied strictly.  
In the decomposition (\ref{decomp}) for each $i$-flow $f_{\mbA,i}$, 
no cycle flow can enter any sink, since it has no departing arcs.
Consequently, the cycle flows may be taken to be zero in (\ref{decomp}) 
without violating (\ref{mccapbnd})--(\ref{mccons}), yielding  
\beq 
f_{\alpha,i} =   
\sum_{1\leq k\leq p_i: \pi_{k,i}\ni\alpha } u_{k,i},  
\label{decompsuff}
\eeq
where $\pi_{1,i},\ldots , \pi_{p_i,i}$ are the $i$-paths and $u_{1,i},\ldots , u_{p_i,i}\geq 0$,
the $i$-path flows. 
In particular,
\beq 
\Hinf [X_i] \stackrel{(\ref{mcdemand})}{=} f_{\out(\sigma_i)}=f_{\into(\tau_i),i} =   
\sum_{k=1}^{p_i} u_{k,i}.  
\label{decompsuff2}
\eeq
%%Recall that the $i$-paths  represent all the  possible acyclic routes through the structure
%%from $\sigma_i$ to $\tau_i$. 

For an arbitrary $\eps>0$,  divide the time axis $\WW$ into epochs of sufficiently
long duration $m\in\NN$ such that
$\forall j\in\NN, i\in [1:n]$, 
\beq
\frac{\mrH \lbr X_i\lp (j-1)m:jm-1\rp \rbr}{m} 
= \frac{\mrH \lbr X_i ( 0:m-1 ) \rbr}{m} \leq \Hinf[X_i] +\eps,
\label{epsmbnd}
\eeq  
where the first equality arises from stationarity.
Next use Huffman coding \cite{cover} to losslessly encode each source-block $ X_i\lp (j-1)m:jm-1\rp$, 
$j\in\NN$, into binary codewords $Z_{i,j}$ of variable length $L_{i,j}$,  
where
\beq
\EX[L_{i,j}] \leq \mrH \lbr X_i ( 0:m-1 ) \rbr + 1  \stackrel{(\ref{epsmbnd})}{\leq} m\Hinf[X_i] + m\eps + 1. 
\label{aveLbnd}
\eeq
Then partition the bits of $Z_{i,j}$ into $p$ consecutive sub-blocks $Z_{i,j,k}$, $k\in[1:p_i]$,
of length $L_{i,j,k}:=\left\lceil \frac {u_{k,i}}{\Hinf [X_i]}L_{i,j}\right\rceil$.
This is always possible since  $\sum_{k=1}^{p_i} L_{i,j,k} \stackrel{(\ref{decompsuff2})}{\geq}  L_{i,j}$,
padding the last sub-blocks with zeros if necessary.

Transmit and route each sub-block $Z_{i,j,k}$ along the $k$-th $i$-path $\pi_{k,i}$.
On every arc $\alpha\in\mbA$ apart from those leaving sources,   
let the arc-signal  be 
$S_\alpha(t)= 0$ when $t\pmod m \neq m-1$ 
and by  
$S_\alpha(t) =\left (Z_{i,j,k}\right )_{i\in[1:n], k\in[1:p_i]: \pi_{k,i}\ni\alpha}$ when $t\equiv jm-1$, $j\in\NN$.\footnote{If an arc is not on any $i$-path,
then its arc signal may be taken to be 0.} 
The arc signals  leaving sources are set to the respective source signals to satisfy  (\ref{Xi}).
Clearly $S$ is setwise causal (Def. \ref{causaldefn}), since every arc-signal is constructed
by  routing  blocks along acyclic paths.
In addition,  $\forall \alpha\in\mbAf$,    
\begin{align}
\Hsup[S_\alpha]  
&=  \frac{1}{m}\mrH\lbr \lp Z_{i,j,k}\rp_{i\in[1:n], k\in[1:p_i]: \pi_{k,i}\ni\alpha}\rbr  
\nn\\
%%\label{Hdetagain}\\
& \leq  \sum_{i\in[1:n], k\in[1:p_i]: \pi_{k,i}\ni\alpha}
\frac{\mrH[ Z_{i,j,k}]}{m}
\label{Hsubadd}\\
&\leq  
 \sum_{i\in[1:n], k\in[1:p_i]: \pi_{k,i}\ni\alpha} \frac{ \EX [ L_{i,j,k}]}{m}
\label{Lijkbnd}\\
&\leq  
  \sum_{i\in[1:n], k\in[1:p_i]: \pi_{k,i}\ni\alpha} 
 \frac{1}{m}+ \frac {u_{k,i}}{m\Hinf [X_i]}\EX [L_{i,j} ]   
\nn\\
&\leq \mrO(1/m) +  
  \sum_{i\in[1:n], k\in[1:p_i]: \pi_{k,i}\ni\alpha} 
 u_{k,i}\frac {m\Hinf[X_i] + m\eps + 1}{m\Hinf [X_i]}   
\nn\\
&= \sum_{i\in[1:n], k\in[1:p_i]: \pi_{k,i}\ni\alpha} 
 u_{k,i} + \mrO(\eps) + \mrO(1/m) 
\nn\\
&\stackrel{(\ref{decompsuff})}{=}   \sum_{i\in[1:n]} f_{\alpha,i}
+ \mrO(\eps) + \mrO(1/m)
\nn\\
&= f_\alpha + \mrO(\eps) + \mrO(1/m) \leq c_\alpha
\nn
\end{align}
for $\eps$ sufficiently small and $m$ sufficiently large.
In the above, the bound (\ref{Hsubadd}) is due to the subadditivity of entropy,
and (\ref{Lijkbnd}) is due to the fact that the expected number of bits 
needed to uniquely specify the value of a random variable is never less than its entropy. 
Furthermore, 
\[
Y_i(jm-1)  
= \lp Z_{i,j,k}\right )_{k=1}^{p_i} \equiv Z_{i,j}
\equiv  X_i\lp (j-1)m:jm-1\rp.
\]
Consequently, $S$ is a solution to the $n$-pairs information network problem $(\Sigma,X,c)$,
establishing achievability (Def. \ref{achdefn}).

\section{Examples}
\label{exsec}

In this section, several examples are given to illustrate the applicability of
Thm. \ref{mainthm}.
However, to begin with  a well-known counterexample is  discussed.

To avoid cluttering the Figures in this section,
 arcs leading out of sources and into sinks are not explicitly
depicted. 

\subsection{Butterfly Network}

\begin{figure}[!t]
\centering
\includegraphics[width=4in]{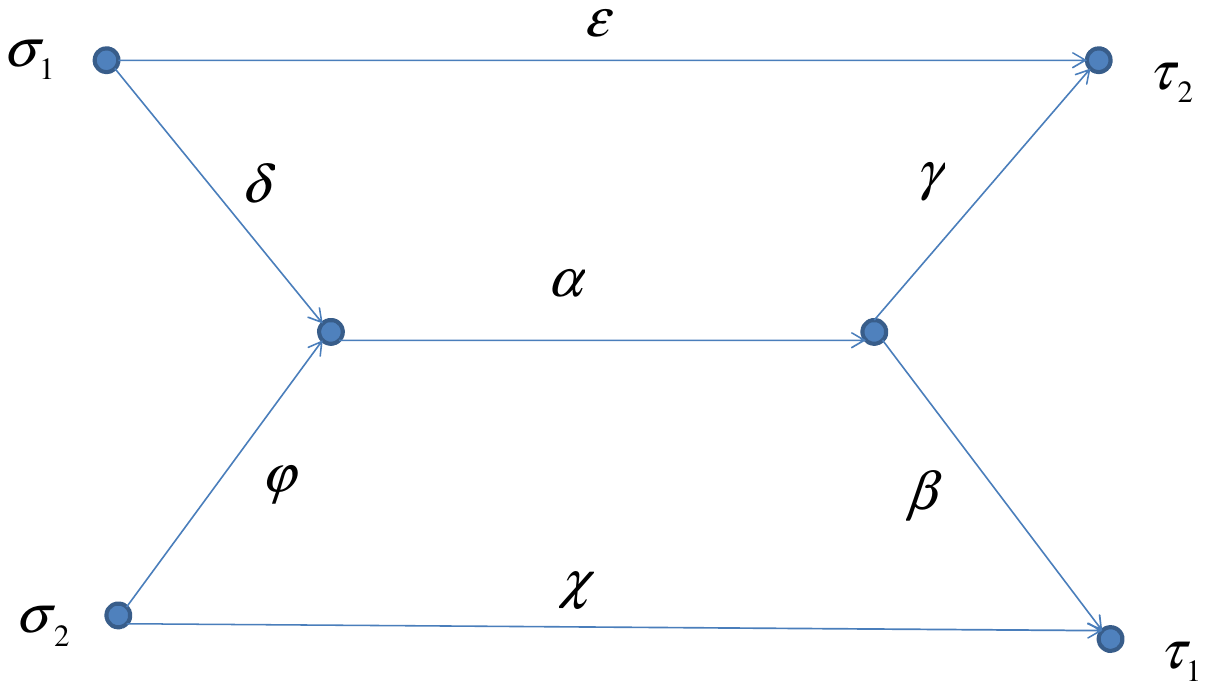}
% where an .eps filename suffix will be assumed under latex, 
% and a .pdf suffix will be assumed for pdflatex; or what has been declared
% via \DeclareGraphicsExtensions.
\caption{Butterfly network}
\label{buttfig} 
\end{figure}

The first example, a 2-pairs butterfly network, is adapted from \cite{yeungTIT95,kramer06}
and depicted in Fig. \ref{buttfig}. For this network it is well-known that
routing does not achieve coding capacity, and it is a useful exercise to verify that it
is not downward dominated.

Consider the viable 2-cut having the set $\mbO^2=\{\alpha\}$ of outgoing arcs in the 2-bundle.
Clearly, both $\beta,\gamma$ are downstream of $\mbO^2=\mbO^2_{*2}$, so 
$\mbC=\{\alpha,\beta,\gamma\}\subseteq\Dom(\mbO^2)$.
No other arcs are downstream of $\mbC$.
Furthermore, the indirect 2-walk concisely represented by $(\varphi,\delta,\eps)$
does not pass through $\alpha$, and neither does the indirect 1-walk
  $(\delta, \varphi,\chi)$. Thus $\mbC$  
is the smallest set satisfying all the conditions of Def. \ref{infodomdefn}),
i.e.  $\mbC=\Dom(\mbO^2)$.
As $\mbC$  does not include any source or sink arcs, 
 this network is not downward dominated (Def. \ref{tridefn}) and Thm. \ref{mainthm}
does not apply.  

\subsection{Examples that Satisfy Lemma \ref{simplestlem}}
\label{cyclesubsec}

\begin{figure}[!t]
\centering
\includegraphics[width=4in]{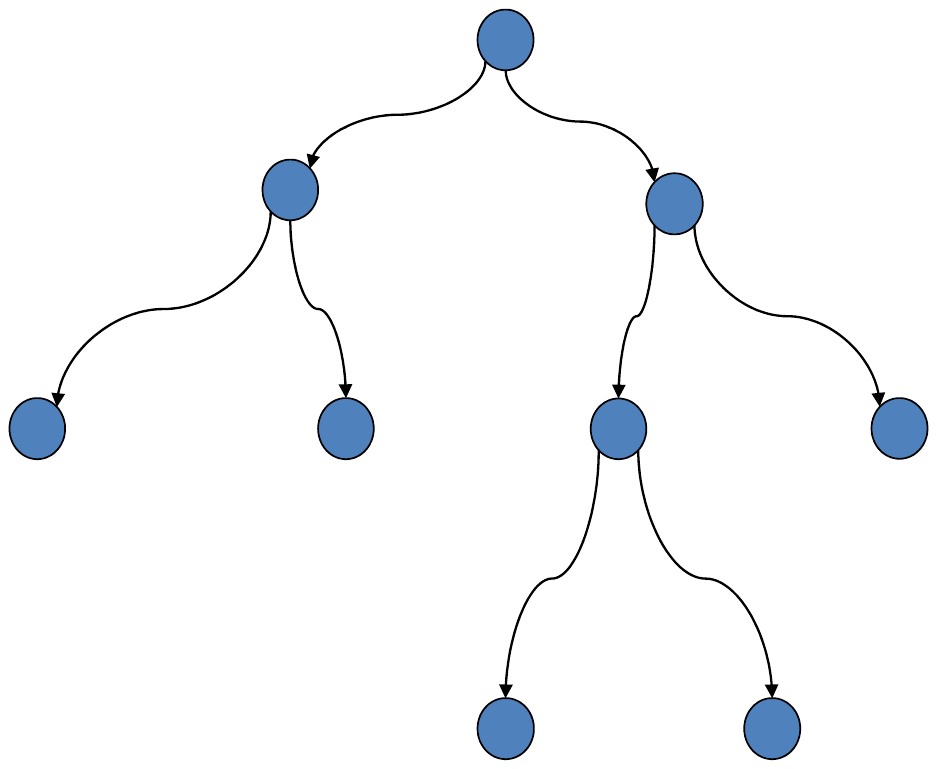}
% where an .eps filename suffix will be assumed under latex, 
% and a .pdf suffix will be assumed for pdflatex; or what has been declared
% via \DeclareGraphicsExtensions.
\caption{A directed tree. Sources and sinks  
may be attached to any of the nodes depicted.}
\label{treefig} 
\end{figure}

\begin{figure}[!t]
\centering
\includegraphics[width=4in]{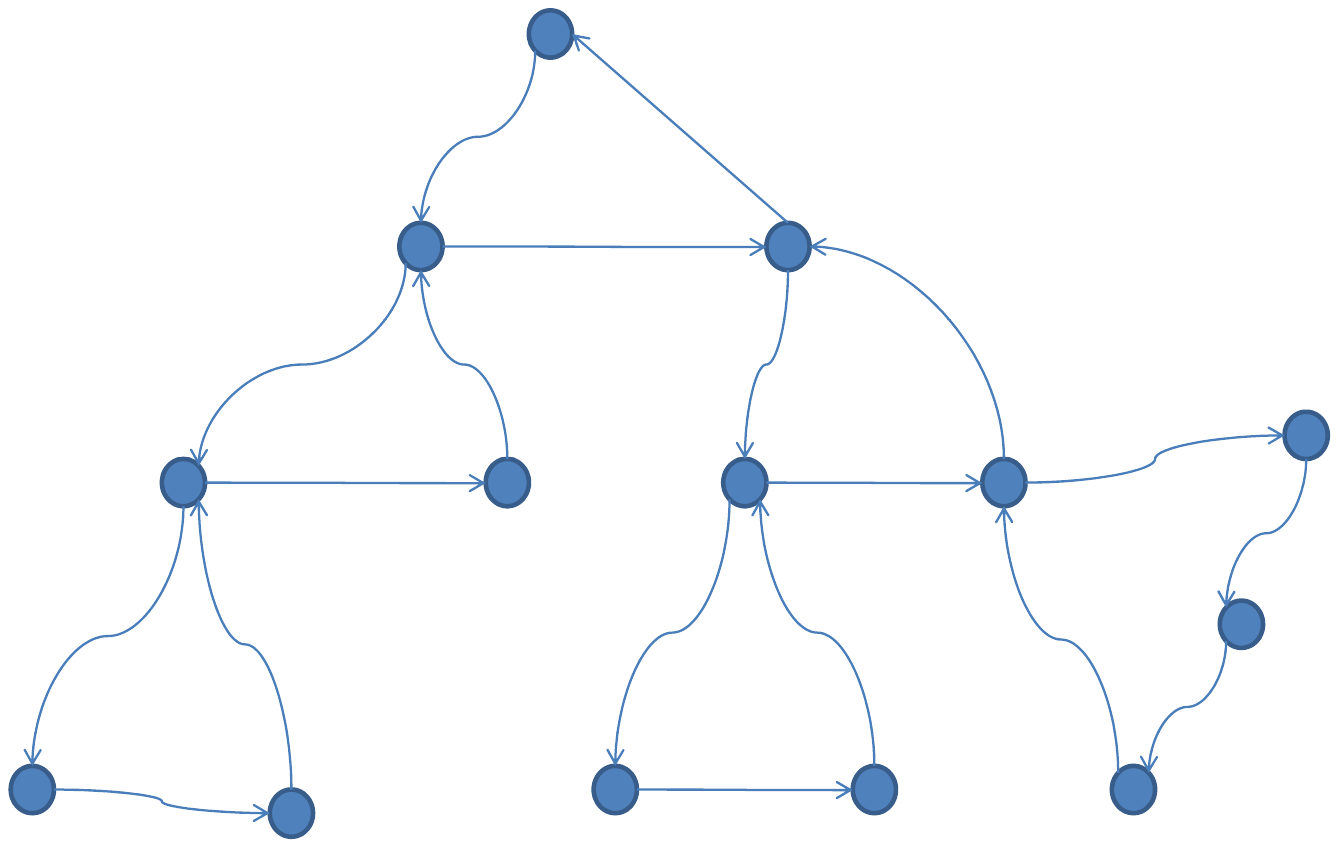}
% where an .eps filename suffix will be assumed under latex, 
% and a .pdf suffix will be assumed for pdflatex; or what has been declared
% via \DeclareGraphicsExtensions.
\caption{A tree of directed cycles. Sources and sinks  
may be attached to any of the nodes depicted.}
\label{cycletreefig} 
\end{figure}

Any network where there is at most one (directed) path from any vertex to any other   
automatically satisfies the conditions of  Lemma \ref{simplestlem},  and is therefore  
dowward dominant and structurally routable (Thm. \ref{mainthm}).
This includes in the  first instance both directed lines  and directed cycles,
agreeing with results in \cite{kramer06,harveyTIT06}.
It also  covers more complicated  structures, 
for instance directed trees (Fig. \ref{treefig}), and 
directed cycles arranged in a line  or tree structure via one or more gateway nodes
(Fig. \ref{cycletreefig}). In all these networks, routing achieves coding capacity regardless
of where sources and sinks are placed.

\begin{figure}[!t]
\centering
\includegraphics[width=4in]{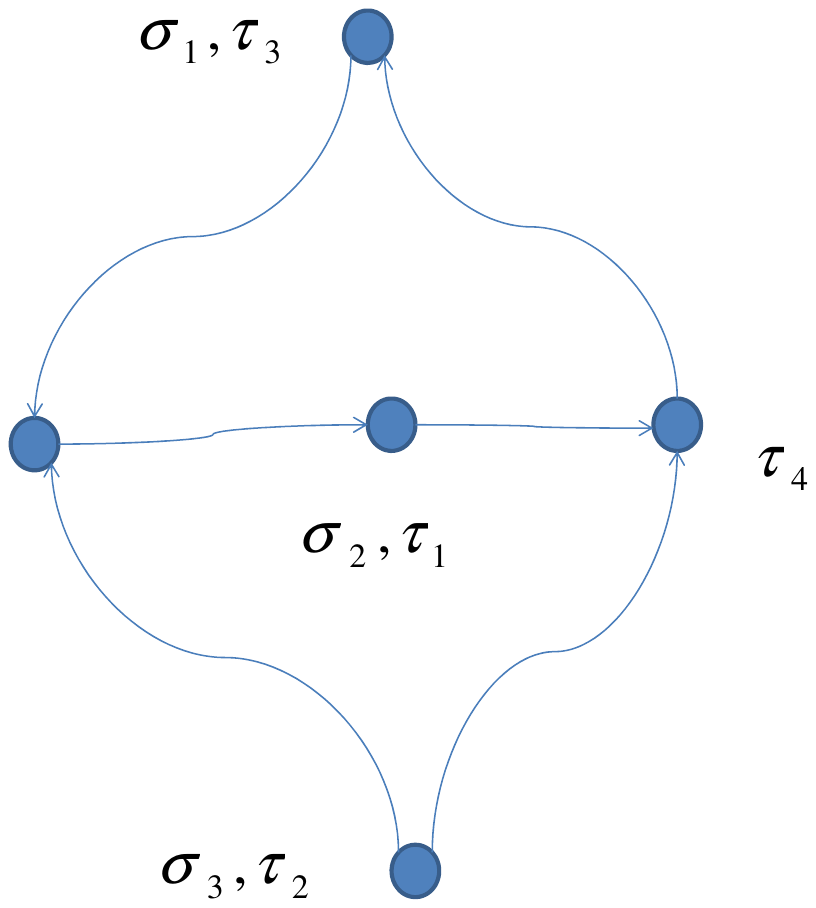}
% where an .eps filename suffix will be assumed under latex, 
% and a .pdf suffix will be assumed for pdflatex; or what has been declared
% via \DeclareGraphicsExtensions.
\caption{A directed version of the Okamura-Seymour Network.  
Only one path exists from any source to any sink.}
\label{okamurafig} 
\end{figure}

In networks where there there are vertex pairs with two or more connecting paths,
 downward dominance will still hold by virtue of Lemma \ref{simplestlem})  
if there is at most one path between each pair of source and sink vertices, 
or at least from each $\sigma_s$ to each $\tau_h$,
where $1\leq s\leq h\leq n$. Examples include directed versions of the undirected Okamura-Seymour
network (Fig. \ref{okamurafig}).   

\subsection{Examples that Satisfy Lemma \ref{simplerlem}} 

\begin{figure}[!t]
\centering
\includegraphics[width=4in]{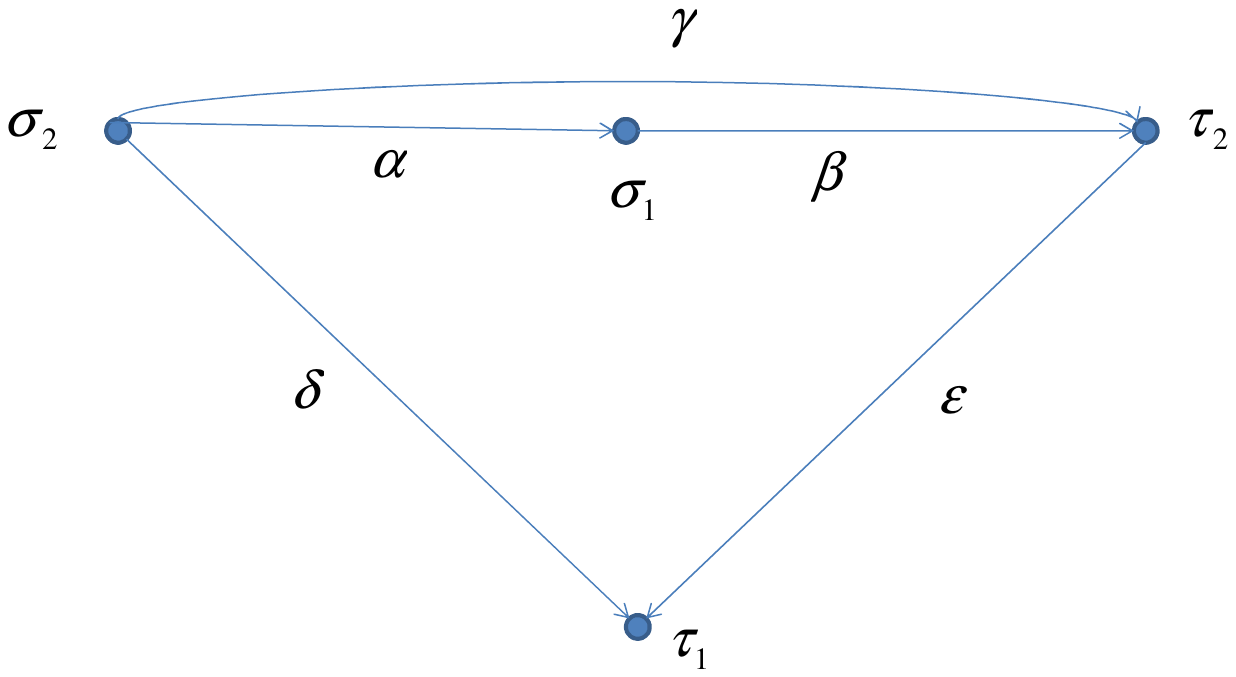}
% where an .eps filename suffix will be assumed under latex, 
% and a .pdf suffix will be assumed for pdflatex; or what has been declared
% via \DeclareGraphicsExtensions.
\caption{An acyclic network covered by Lemma \ref{simplerlem}.}
\label{simplerlemfig} 
\end{figure}

Now consider the acyclic 2-pairs network in Fig. \ref{simplerlemfig}.
Observe that there is one 1-path, 
concisely represented by the arc-sequence $\beta\eps$, 
but two 2-paths,  $\alpha\beta$ and $\gamma$.
Hence  Lemma \ref{simplestlem} cannot be applied. 
Neither 
 would it become applicable   
if the indices 1 and 2 were relabelled 2' and 1' respectively. 
To see this, consider the viable 2'-cut with $\mbO^{2'}=\{\beta\}$.
Clearly $\mbA^{2'}\cap\mbO^{2'}$ and 
$\mbA^{1'}\cap\mbO^{2'}\neq\emptyset$, 
since the 1'-path $\alpha\beta$ and 2'-path $\beta\eps$ both pass through $\mbO^{2'}$.
However, the path $\gamma$ from $\sigma_1'$ to $\tau_1'$ does not. 

In this instance, Lemma \ref{simplerlem} can be applied.
The possible viable 2-cuts have sets $\mbO^2$ of outgoing arcs in the 2-bundle 
 equal to either  $\{\alpha,\gamma\}$ or $\{\beta,\gamma\}$.
In the  first case, $\mbO^2$ has no intersection with any arcs in the 1-bundle,
and all 2-paths obviously pass through it.
In the second case, all paths from $\sigma_s$ to $\sigma_h$, $1\leq s\leq h\leq 2$,
pass through $\mbO^2$. This the requirements of the Lemma are met and the network is downward dominant. 

\begin{figure}[!t]
\centering
\includegraphics[width=4in]{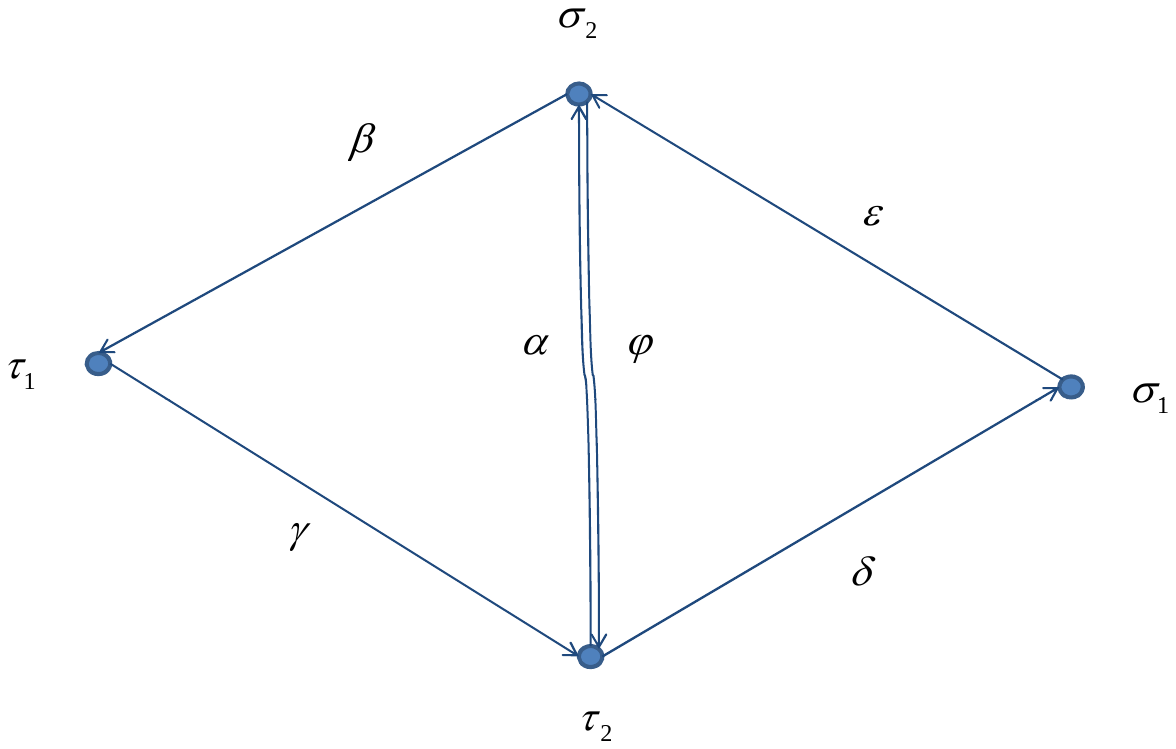}
% where an .eps filename suffix will be assumed under latex, 
% and a .pdf suffix will be assumed for pdflatex; or what has been declared
% via \DeclareGraphicsExtensions.
\caption{A cyclic network covered by Lemma \ref{simplerlem}.}
\label{cycfig} 
\end{figure}

Another example of the use of Lemma \ref{simplerlem}) is the the cyclic 2-pairs network of Fig. \ref{cycfig}.
Observe that there is one 1-path, $\eps\beta$ 
and two 2-paths, $\varphi$ and $\beta\gamma$.  
The possible viable 2-cuts have sets $\mbO^2$ of outgoing arcs in the 2-bundle 
 equal to either  $\{\varphi,\beta\}$ or $\{\varphi,\gamma\}$. 
In the  second case, $\mbO^2$ has no intersection with any arc in the 1-bundle,
and all 2-paths obviously pass through it.
In the first case,  $\mbO^2$ intersects all 2-paths and a 1-path, $\eps\beta$,
and it can be seen that all from $\sigma_s$ to $\sigma_h$, $1\leq s\leq h\leq 2$,
pass through $\mbO^2$. This the requirements of the Lemma are met and the network is downward dominant. 

\section{Conclusion}
\label{concsec}

This chapter examined the routability of possibly cyclic $n$-pairs information networks from a structural perspective.
The concept of
downward dominance was introduced, and it was shown that for networks with downward dominated structures,  
routability and achievability are  equivalent, i.e. 
  a given combination of source signals, demand rates and channel capacities is achievable 
iff the network supports a feasible multicommodity flow.

Downward dominance is a conservative structural condition, and future work will focus on trying to relax it.
The inductive nature of the proof of necessity here requires it directly,
so any generalisation may need a very different analysis technique.   
 
%to the examples in the literature \cite{yanTIT06,harveyTIT06}
%of special network structures 
%for which routing always achieves capacity.  

\begin{acknowledgement}
The author acknowledges discussions  
on the  decentralised control version of this problem 
with Prof. Rob Evans at the University of Melbourne.
He is also indebted to the anonymous ISIT11 reviewer 
who pointed out an error in the original version of (\ref{flowdemand}).
\end{acknowledgement}

\bibliographystyle{spmpsci}
\bibliography{rep1ref}

%%\bibliographystyle{spmpsci}
%%\bibliography{rep1ref}

\end{document}